\documentclass[sigplan,screen,nonacm]{acmart}
\usepackage{tabularx}
\usepackage{booktabs}
\usepackage{listings}
\renewcommand\footnotetextcopyrightpermission[1]{}

\title[What Field Enumeration Costs to Compile in C++26]{What Field Enumeration Costs to Compile in C++26: Native Reflection, Structured-Binding Packs, Reflection Libraries and Hand-Written Serializers on GCC 16}
\author{Denys Kachalenko}
\orcid{0009-0003-8597-329X}
\affiliation{\institution{Independent researcher}\country{Ukraine}}
\email{denys.kachalenko@gmail.com}

\begin{document}

\begin{abstract}
C++26 adds standard ways to enumerate a struct's fields, among them static reflection (P2996, with P3394 annotations) and structured-binding packs (P1061), alongside hand-written code, X-macros and libraries (Boost.PFR, reflect-cpp, Glaze) that now ship pre-C++26 and C++26 paths. We time each in a struct serializer with byte-identical output on GCC 16.2.0, at the front end and at \texttt{-O2}, in randomized interleaved rounds with code-level checks of every single-factor contrast. At the front end, inline native enumeration in the form used in several of P2996's own examples, an expansion statement over the \texttt{std::\allowbreak{}span} of \texttt{define\_\allowbreak{}static\_\allowbreak{}array}, costs about as much per struct of 16 scalar fields as Boost.PFR and 7.96$\times$ a P1061 pack, itself about as cheap as hand-written code. The same loop over a constant array costs 0.47$\times$ as much: about half of the span form's cost comes with the span. The ratio to P1061 shrinks on a heterogeneous type graph and at \texttt{-O2}. In a whole translation unit at the front end the library outweighs the mechanism: 200 structs take 0.59 s with inline native enumeration, 0.56 s with Boost.PFR and 3.86--4.13 s with the C++26 paths of reflect-cpp and Glaze, whose visitors also compute field names. Per type, each library's C++26 path is cheaper than its pre-C++26 path at 16 fields, but Glaze's gain reverses at 128 fields (pinned release v9.1.0); Boost.PFR's C++17 path grows almost quadratically with the number of types. Hand-written serializers sharing one overloaded name grow superlinearly with the number of types. The \texttt{-freflection} flag alone makes reflection-free code up to 1.264$\times$ costlier per type on GCC 16.2.0 and up to 2.578$\times$ on 16.1.0, so comparisons between mechanisms depend on the point release. Part of the span form's cost grows with the number of structs, which we trace to GCC registering block-scope statics in quadratic time (reported as GCC PR128928).
\end{abstract}

\maketitle

\section{Introduction}\label{sec:1}

Serialization is boilerplate: for every type, code that visits every field. C++ has attacked it with the preprocessor (X-macros), with library ``reflection'' built on aggregate-initialization probing, structured bindings and parsing of compiler-generated function names (Boost.PFR, reflect-cpp, Glaze), and now with C++26 language features: static reflection (P2996) with annotations (P3394), structured bindings that introduce a pack (P1061), and expansion statements (P1306), which iterate over reflected members and can also destructure a struct directly. For a practitioner the question is concrete: \emph{what does each of these cost to compile, and what does each save in source?}

Native P2996 compile time has only begun to be measured. Romeo~\cite{romeo2026a} times native reflection alone on single translation units; mp-units~\cite{pusz2026} runs an in-repo A/B whose author considers the sign of the effect reliable but not its magnitude; simdjson reports, on clang-p2996, a 6\% slowdown of its whole build with reflection enabled~\cite{lemire2025cppcon}; and the field-reflection library~\cite{matsuda2026} lists in its README the per-type cost of its \texttt{std::\allowbreak{}meta} backend next to Boost.PFR's, as best-of-three numbers. We found no published measurement, as of November 2026, that compares native reflection with each library's two paths, with P1061, or with hand-written and macro baselines on one serialization task (\S{}5). A tooling vendor's blog~\cite{wholetomato2026} calls compile-time impact ``the biggest open question''.

\textbf{Research questions.}

\begin{itemize}
\item \textbf{RQ1 --- mechanisms.} Per type, what does field enumeration cost to compile with P2996, with P1061, through a library's callback API, and by hand; and what does a P3394 annotation query add?
\item \textbf{RQ2 --- library backends.} Each of Boost.PFR, reflect-cpp and Glaze can be built on its pre-C++26 path or its C++26 path from the same source. Does switching change its compile cost, and how does that change scale with the number of fields per struct?
\item \textbf{RQ3 --- per-type serializer idioms.} Which documented customization idioms make hand-written serialization grow faster than linearly with the number of types, and why?
\item \textbf{RQ4 --- what the numbers depend on.} How large are the fixed costs (\texttt{\textless{}meta\textgreater{}}, library headers, the \texttt{-freflection} flag itself), how do the results change at \texttt{-O2}, on richer corpora and a heterogeneous type graph, between sessions, and between GCC 16.1.0 and 16.2.0?
\item \textbf{RQ5 --- maintainability.} How much per-type source does each approach need?
\end{itemize}

\textbf{Contributions.}

\begin{enumerate}
\item A head-to-head compile-time comparison of field enumeration for serialization: P2996, P1061, hand-written code, X-macros and three libraries, each library on both of its paths (three same-source backend A/Bs), on GCC 16.2.0, with key arms (configurations, \S{}3.1) also compiled by GCC 16.1.0 in the same rounds.
\item A factorial decomposition of native and library enumeration cost: enumeration mechanism (P2996 or P1061) $\times$ API shape (inline or callback), the annotation query, and the library machinery around the same primitive. Native enumeration here means the form timed throughout, an expansion statement over the \texttt{std::\allowbreak{}span} of \texttt{define\_\allowbreak{}static\_\allowbreak{}array} (Listing~\ref{lis:1}); the decomposition shows that, at the front end, this form matches Boost.PFR only when written inline, not behind the same callback API. A supplementary session times four formulations of the same member query. On GCC 16.2.0, about half of the span form's front-end cost comes with the span, which we attribute mainly to GCC's evaluation of the iterating expansion statement that C++26 specifies over it. Over a constant array, inline native enumeration costs 0.49$\times$ Boost.PFR's callback \texttt{for\_\allowbreak{}each\_\allowbreak{}field}.
\item A comparison of ten customization idioms for per-type serializers, with the number of call sites and the number of candidates varied separately; serializers that share one overloaded name grow superlinearly with the number of types.
\item Controls that bound the measurement itself: a code-identical pair that differs only in what a library flag adds to the parse, the cost of the \texttt{-freflection} flag on both compiler versions, the overhead of \texttt{-ftime-report}, a value-level correctness oracle with a negative control, and code-level manipulation checks on every pair of arms meant to differ in a single factor.
\item An open artifact: generator, harness, raw per-run data and environment records.
\end{enumerate}

\section{Background}\label{sec:2}

\textbf{Enumerating fields in C++26.} P2996 reflects an entity with \texttt{\textasciicircum{}\textasciicircum{}} to a \texttt{std::\allowbreak{}meta::\allowbreak{}info}; \texttt{std::\allowbreak{}meta::\allowbreak{}nonstatic\_\allowbreak{}data\_\allowbreak{}members\_\allowbreak{}of} returns a type's fields, and a splice \texttt{v.\allowbreak{}[:m:]} accesses one. Our native serializers iterate those fields with an expansion statement (\texttt{template for}, P1306~\cite{wg21p1306}) over \texttt{std::\allowbreak{}define\_\allowbreak{}static\_\allowbreak{}array} (P3491~\cite{wg21p3491}) (Listing~\ref{lis:1}). This is one of several ways to write native enumeration; every job of S1 and S2 with a native arm times it, and a supplementary session times three others (\S{}4.2). P3394 attaches values to declarations (\texttt{[[=\allowbreak{}RB::\allowbreak{}FTransient\{\}]]}), which reflection can query, e.g. to skip a field. P1061 lets a structured binding introduce a pack: inside a template, \texttt{const auto\& [... m] =\allowbreak{} v;} binds every field of a struct such as ours (all non-static data members accessible and declared in one class), so a fold expression visits them all --- enumeration without names or annotations (P1061R10~\cite{wg21p1061} allows such a pack only in a templated entity). An expansion statement can also destructure the struct itself, \texttt{template for (\allowbreak{}const auto\& f : v)}, with the same limits (no names, no annotations) but outside templates too; we did not time this form. GCC 16 implements P2996R13~\cite{wg21p2996}, P3394R4~\cite{wg21p3394}, P3491R3, P1306R5 and P1061R10; reflection needs \texttt{-freflection}, which GCC labels experimental.

\textbf{Pre-C++26 library techniques.} Libraries count an aggregate's fields by probing aggregate initialization with an object convertible to any type, bind them through a generated ladder of structured bindings (one case per field count), and recover names by parsing \texttt{\_\allowbreak{}\_\allowbreak{}PRETTY\_\allowbreak{}FUNCTION\_\allowbreak{}\_\allowbreak{}} or \texttt{std::\allowbreak{}source\_\allowbreak{}location::\allowbreak{}function\_\allowbreak{}name(\allowbreak{})} of a template instantiated on the address of a member of an \texttt{extern} object that is never defined. \textbf{Boost.PFR}'s classic path~\cite{boostpfr} works this way; under \texttt{-std=\allowbreak{}c++26} on GCC 16 it switches by itself to a path built on P1061 packs. Its P2996 engine (\texttt{BOOST\_\allowbreak{}PFR\_\allowbreak{}USE\_\allowbreak{}CPP26\_\allowbreak{}REFLECTION}) is not reached by \texttt{for\_\allowbreak{}each\_\allowbreak{}field} (\S{}3.3), and on its own it cannot serialize runtime objects at the pinned release, because its member access (\texttt{detail::\allowbreak{}reference\_\allowbreak{}by\_\allowbreak{}index}) is \texttt{consteval}. \textbf{reflect-cpp}~\cite{reflectcpp} has a C++20 path and, since v0.26.0 (2026-10-18), an opt-in C++26 path; its documentation states that the latter is ``considerably cheaper'' to compile, without measurements~\cite{reflectcppdocs}. \textbf{Glaze}~\cite{glaze} has a traditional backend and a P2996 backend. At the pinned release the traditional backend counts members with a recursive chain of aggregate-initialization probes (\texttt{V\{Args\{\}...,\allowbreak{} any\_\allowbreak{}t\{\}\}}), binds them through a generated ladder of structured bindings, parses names from \texttt{\_\allowbreak{}\_\allowbreak{}PRETTY\_\allowbreak{}FUNCTION\_\allowbreak{}\_\allowbreak{}} and accepts at most 128 members per struct; Glaze's documentation calls the count a structured-binding probe~\cite{glazedocs}. Glaze selects its P2996 backend by itself when the compiler defines \texttt{\_\allowbreak{}\_\allowbreak{}cpp\_\allowbreak{}impl\_\allowbreak{}reflection}, as GCC 16 does under \texttt{-freflection}.

\textbf{Per-type customization idioms.} Many serialization and formatting libraries let users write one function per type under one name and find it by name lookup: nlohmann::json's \texttt{to\_\allowbreak{}json} and its \texttt{NLOHMANN\_\allowbreak{}DEFINE\_\allowbreak{}TYPE\_\allowbreak{}NON\_\allowbreak{}INTRUSIVE} macro, Boost.JSON's \texttt{tag\_\allowbreak{}invoke}, \texttt{BOOST\_\allowbreak{}DESCRIBE\_\allowbreak{}STRUCT}, \{fmt\}'s \texttt{format\_\allowbreak{}as}, cereal's external \texttt{serialize}, and Boost.Serialization's non-intrusive \texttt{serialize}. When N user types live in one namespace, each resolution of such a name considers N candidates. nlohmann::json, Boost.JSON and \{fmt\} make the unqualified call inside their own dispatch template, so it is resolved once per serialized type; code that calls the name directly resolves it once per call site. Hidden friends (nlohmann's \texttt{\_\allowbreak{}INTRUSIVE} macro, \texttt{BOOST\_\allowbreak{}DESCRIBE\_\allowbreak{}CLASS}), member functions (cereal's preferred form, protobuf) and class-template specializations (\texttt{glz::\allowbreak{}meta}, \texttt{fmt::\allowbreak{}formatter}) do not create that set.

\section{Methodology}\label{sec:3}

\subsection{Arms}\label{sec:3.1}

Every arm writes the same little-endian binary format through one shared value encoder, so outputs are byte-identical; arms differ only in how they enumerate fields.

\begin{table*}[t]
\small
\caption{The serializer arms: how each enumerates the fields, and the source it needs per type.}
\label{tab:arms}
\begin{tabularx}{\textwidth}{>{\raggedright\arraybackslash}X>{\raggedright\arraybackslash}Xl}
\toprule
Arm & Field enumeration & Per-type source \\
\midrule
\texttt{manual} / \texttt{manual\_\allowbreak{}unique} & hand-written; one overloaded name / a distinct name per type & every field \\
\texttt{xmacro} / \texttt{xmacro\_\allowbreak{}unique} & X-macro field list; one overloaded name / distinct names & every field \\
\texttt{reflect} & P2996, plus a P3394 annotation query per member & none \\
\texttt{reflect\_\allowbreak{}plain} & P2996 & none \\
\texttt{reflect\_\allowbreak{}visit} & P2996 behind a callback API shaped like \texttt{boost::\allowbreak{}pfr::\allowbreak{}for\_\allowbreak{}each\_\allowbreak{}field} & none \\
\texttt{reflect\_\allowbreak{}array} / \texttt{reflect\_\allowbreak{}index} / \texttt{reflect\_\allowbreak{}seq} (S3 only) & P2996, the same member query: the expansion statement over a constant array (\texttt{[:reflect\_\allowbreak{}constant\_\allowbreak{}array(\allowbreak{}...):]}) / a P1061 index pack bound from \texttt{make\_\allowbreak{}index\_\allowbreak{}sequence} (reflect-cpp's C++26 form) / an index-sequence helper & none \\
\texttt{p1061} / \texttt{p1061\_\allowbreak{}visit} & P1061 pack, inline / behind the same callback API & none \\
\texttt{pfr} & Boost.PFR, default under \texttt{-std=\allowbreak{}c++26} (its P1061 path) & none \\
\texttt{pfr\_\allowbreak{}classic} & Boost.PFR, C++17 path (\texttt{-DBOOST\_\allowbreak{}PFR\_\allowbreak{}USE\_\allowbreak{}CPP26=\allowbreak{}0}) & none \\
\texttt{pfr\_\allowbreak{}native} & Boost.PFR with \texttt{-DBOOST\_\allowbreak{}PFR\_\allowbreak{}USE\_\allowbreak{}CPP26\_\allowbreak{}REFLECTION=\allowbreak{}1}: compiles to the same code as \texttt{pfr} (identical disassembly, \S{}3.3); the flag only adds headers to the parse (code-identical control, \S{}4.2) & none \\
\texttt{rflcpp} / \texttt{rflcpp\_\allowbreak{}native} & reflect-cpp, C++20 path / C++26 path & none \\
\texttt{glaze\_\allowbreak{}classic} / \texttt{glaze\_\allowbreak{}native} & Glaze, traditional backend (forced) / P2996 backend & none \\
\bottomrule
\end{tabularx}
\end{table*}

We call the arms of S1 and S2 other than those of reflect-cpp and Glaze \emph{light}; the three S3 arms run only in FrontendIdiom. The four arms of those two libraries, which we call \emph{heavy}, cost 12.75 to 30.56 ms per struct at the front end (Table~\ref{tab:2}), so the jobs that time them stop at N = 200 or earlier. In the job table of \S{}3.5 and in \S{}4.1, \S{}4.5 and \S{}4.6, ``the libraries'' means these heavy arms; the Boost.PFR arms are light.

\begin{lstlisting}[language=C++, morekeywords={constexpr,consteval,requires}, float=*t, label=lis:1, caption={The core of the arms (from \texttt{Include/\allowbreak{}RB} and the generator). \texttt{Enc::\allowbreak{}Any} is the shared value encoder; it hands nested structs back to the arm through \texttt{FSelf}.}]
// reflect_plain: P2996 queries, P1306 expansion statement, P3491 static array
template<typename Type>
void Write(FByteWriter& Writer, const Type& V)
{
    constexpr auto Context = std::meta::access_context::unchecked();
    template for (constexpr auto Member :
        std::define_static_array(std::meta::nonstatic_data_members_of(^^Type, Context)))
    {
        Enc::Any<FSelf>(Writer, V.[:Member:]);
    }
}

// reflect_array (S3): the same loop over a constant array instead of the std::span
//     template for (constexpr auto Member :
//         [:std::meta::reflect_constant_array(std::meta::nonstatic_data_members_of(^^Type, Context)):])

// p1061: structured-binding pack and fold expression
template<typename Type>
void Write(FByteWriter& Writer, const Type& V)
{
    const auto& [... Members] = V;
    (Enc::Any<FSelf>(Writer, Members), ...);
}

// manual_unique: generated, one non-template function per type
inline void WriteManualS0(RB::FByteWriter& Writer, const S0& V)
{
    RB::Enc::Value(Writer, V.M0);
    // ... one statement per field
}
\end{lstlisting}

The callback arms (\texttt{reflect\_\allowbreak{}visit}, \texttt{p1061\_\allowbreak{}visit}) move the same loop or fold into a function template \texttt{ForEachField(\allowbreak{}V,\allowbreak{} Func)} and pass it \texttt{[\&](\allowbreak{}const auto\& Field) \{ Enc::\allowbreak{}Any\textless{}FSelf\textgreater{}(\allowbreak{}Writer,\allowbreak{} Field); \}}, the lambda that the \texttt{pfr} arm passes to \texttt{boost::\allowbreak{}pfr::\allowbreak{}for\_\allowbreak{}each\_\allowbreak{}field}. \texttt{reflect} wraps the call in \texttt{if constexpr (\allowbreak{}!IsTransient(\allowbreak{}Member))}, where \texttt{IsTransient} tests \texttt{annotations\_\allowbreak{}of\_\allowbreak{}with\_\allowbreak{}type(\allowbreak{}Member,\allowbreak{} \textasciicircum{}\textasciicircum{}RB::\allowbreak{}FTransient)}.

Library arms call the libraries' field visitors (\texttt{boost::\allowbreak{}pfr::\allowbreak{}for\_\allowbreak{}each\_\allowbreak{}field}, \texttt{rfl::\allowbreak{}to\_\allowbreak{}view(\allowbreak{}v).\allowbreak{}apply}, \texttt{glz::\allowbreak{}for\_\allowbreak{}each\_\allowbreak{}field}), not their own serializers, so every arm emits the same bytes; each includes the narrowest library header that provides its visitor. Nested structs, also inside vectors and optionals, go back to the arm's own enumeration. Further arms serve as controls: \texttt{*\_\allowbreak{}split} (one function per call site, for \texttt{-O2}), \texttt{*\_\allowbreak{}noflag} (compiled with \texttt{-fno-reflection}), \texttt{*\_\allowbreak{}noreport} (compiled without \texttt{-ftime-report}), \texttt{arm@GCC161} (the same arm compiled by GCC 16.1.0 in the same rounds), and ten overload-matrix variants (\S{}4.4).

\subsection{Corpora}\label{sec:3.2}

A generator emits N structs and one benchmark translation unit per arm that serializes each struct once. \emph{Scalar}: M fields per struct, cyclically from eight integer, floating-point and \texttt{bool} types; every struct has the same member list. \emph{Rich}: the same with strings, vectors and optionals. \emph{Hetero}: a type graph whose field counts vary per struct (log-normal, median 8, at most 32 at N = 1,600), with enums (10\% of fields), nested structs (14\%), and vectors (8\%) and optionals (6\%) of structs, at most six levels deep. It is prefix-stable --- the first k types are the same for every N --- so the marginal cost of added types is well defined. Every benchmark file serializes objects declared \texttt{extern}, so at \texttt{-O2} the optimizer cannot fold known field values into the generated code.

\subsection{Correctness and manipulation checks}\label{sec:3.3}

\textbf{Value-level oracle.} Integer, floating-point and string fields receive values that are distinct and non-zero per field; bool fields alternate; strings and vectors are non-empty and optionals mixed. Each arm's output must equal the hand-written reference byte for byte, and a serializer that writes the fields in reverse order must be rejected --- otherwise the oracle reports itself blind. The oracle is built for every flag set a timed arm uses (reflection-free arms without \texttt{-freflection}), and each build must report the expected library-backend configuration; it runs on the scalar and rich corpora at M = 16 and M = 128 and on the heterogeneous corpus at the largest N timed (1,600 types, on a sample that includes every shape class). Further oracles cover the ten overload variants and the P3394 path (a \texttt{Transient} member must be skipped). A CTest suite runs every arm that can nest (all but \texttt{xmacro\_\allowbreak{}unique}) on a nested record with containers, under each library backend.

\textbf{Manipulation checks.} For every pair of arms meant to differ in a single factor (the backend A/Bs, the annotation query, API shape, mechanism, the native formulations of S3, Boost.PFR's machinery, overloaded against distinct names, the flag-only controls and the split controls), the harness compiles both arms and checks, in three ways, that the intended difference is present on one side only. In the object files, a marker function must exist on one side only for each backend A/B (for example \texttt{glz::\allowbreak{}detail::\allowbreak{}get\_\allowbreak{}member\_\allowbreak{}ref} only in the P2996 build of Glaze, \texttt{tie\_\allowbreak{}as\_\allowbreak{}tuple} only in Boost.PFR's C++17 path, \texttt{rfl::\allowbreak{}internal::\allowbreak{}cpp26::\allowbreak{}bind\_\allowbreak{}to\_\allowbreak{}tuple} only in reflect-cpp's C++26 path). In the preprocessed source, the intended difference must appear on one side only (the annotation query only in \texttt{reflect}; the \texttt{nonstatic\_\allowbreak{}data\_\allowbreak{}members\_\allowbreak{}of} call only in the P2996 arm and the structured-binding pack only in the P1061 arm). A pair meant to be identical --- \texttt{pfr}/\texttt{pfr\_\allowbreak{}native} and every \texttt{-fno-reflection} control --- must have the same disassembly. Every arm's flags must reach the exact compiler command that is timed, and \texttt{-fno-reflection} must remove \texttt{\_\allowbreak{}\_\allowbreak{}cpp\_\allowbreak{}impl\_\allowbreak{}reflection}. The checks for a differing pair do not exclude other differences: the P2996 and P1061 arms also differ in parsing \texttt{\textless{}meta\textgreater{}} and in the expansion statement, and the span and array forms of S3 in the calls that \texttt{define\_\allowbreak{}static\_\allowbreak{}array} adds (\S{}4.2, \S{}6). These checks exist because a flag documented as a backend switch can fail to reach the code being called: \texttt{BOOST\_\allowbreak{}PFR\_\allowbreak{}USE\_\allowbreak{}CPP26\_\allowbreak{}REFLECTION=\allowbreak{}1} changes nothing in \texttt{boost::\allowbreak{}pfr::\allowbreak{}for\_\allowbreak{}each\_\allowbreak{}field} under \texttt{-std=\allowbreak{}c++26} on GCC 16 --- the object files define the same functions, with identical disassembly, with and without it. Pairs that differ in many factors at once (native against Boost.PFR, a library against the native arm) have no single expected difference and are checked only through the oracle. The checks ran once for each compiler and are bound to the measurement code (a fingerprint of the harness modules, \texttt{Include/\allowbreak{}RB} and the library pins) and to each compiler's \texttt{cc1plus} hash: a session starts only when each of its compilers has a passing check for exactly that code: S1 and S2 had one on both compilers, and S3, which uses GCC 16.2.0 only, had one for its own code, which adds the three arms. Before each session the library pins are verified again and the CTest suite is re-run with GCC 16.2.0.

\textbf{Code provenance.} The code of this study was written with an AI assistant under the author's direction (Acknowledgments); it is verified by the checks above and by regenerating every number in the paper from the archived raw data and checking it against the text (Artifact).

\subsection{What is measured}\label{sec:3.4}

The metric is the compiler's own \texttt{-ftime-report} TOTAL: wall time measured inside the compiler proper (GCC 16 reports wall time and garbage-collected memory, not CPU time), together with its phase and sub-phase timers and its garbage-collected (GGC) memory. For every compile the harness also stores the external wall time and, read from the exited process, its user and kernel CPU time, CPU cycles, peak working set, peak commit and page faults. The harness runs \texttt{cc1plus} directly with exactly the command line the \texttt{g++} driver generates (\texttt{-\#\#\#}), at high process priority, restricted to the performance cores of the hybrid CPU. Running \texttt{cc1plus} directly matters: a child of a high-priority driver runs at normal priority on Windows, so launching through the driver would leave the timed process unprotected. Two modes: \texttt{-fsyntax-only} (front end: parsing, template instantiation, constant evaluation) and \texttt{-O2 -c} (adds optimization and code generation; the assembly is written to the null device, so the assembler is not run). Per-type numbers exclude fixed costs: they are slopes over N (\S{}3.6) or marginal costs, the cost at N structs minus the cost at N = 0.

\subsection{Protocol}\label{sec:3.5}

A session runs the jobs below. In each job, one warm-up round and eleven measured rounds each compile every cell (arm $\times$ sweep value) once, in a seeded random order, between two compiles of a fixed sentinel cell that tracks drift. Arms compared in a ratio always come from the same job and so share rounds. Before timing, the harness records the machine (CPU and core layout, power plan, antivirus state, per-CPU load and busiest processes), every compiler's hashes and configure line, the library pins and a fingerprint of the measurement code, of which it keeps a copy; it binds the session to its compilers, code and protocol, so a resumed session cannot silently mix them. It runs under the High performance power plan. Rounds are not repeated: the machine's CPU load during each compile is recorded with that compile, and every estimate is a median or a fit over the eleven measured rounds. Two sessions on different days measure the full plan on GCC 16.2.0; the comparison jobs compile their arms with GCC 16.2.0 and GCC 16.1.0 side by side in the same rounds. A third, short session (S3, 2026-11-21) ran one supplementary job, FrontendIdiom, on GCC 16.2.0 with the same protocol; it was added after the review of the first two to time other native formulations, and its preflight checked the three new arms as it checks all others. Before the session, six of its seven arms (all but the array form, added later) were compiled once each informally and, with the array form added, a one-round smoke test of the whole job ran; neither is used here. Its analysis rules were fixed before the session: formulations are compared only within the job, an effect is attributed to the expansion statement's lowering only through the array/span pair, and the arms shared with S1 are compared with S1 descriptively. Those shared arms allocate the same GGC memory as in S1 at every N.

S1 and S2 each ran 8,484 compiles, none of which failed. During the measured compiles the machine's total CPU load (over all 20 logical CPUs, of which the single-threaded compiler alone accounts for up to 5\%) had a median of 8.0\% in S1 and 7.3\% in S2, and exceeded 15\% for 5.1\% and 2.0\% of compiles. Dropping those compiles before taking the cell medians changes 90 of S1's 96 per-arm slopes that pass the linearity check and rise by at least ten timer steps (in every job but ReportOverhead, which is compared on external wall time) by no more than 1\% (the largest change is 2.1\%), and all 97 of S2 by no more than 0.6\%. For compiles longer than 2 s, TOTAL lies within 3\% of the process's CPU time (user + kernel) in 99.3\% of cases in S1 and 99.2\% in S2, so these compiles spent almost no time waiting off the CPU; the sentinel cell stayed within 4.7\% (S1) and 5.2\% (S2) of its median in every job. S3 ran 444 compiles, none of which failed, at a median load of 8.0\% (2.9\% of compiles above 15\%). Dropping its compiles above 15\% load changes none of its slopes by more than 0.3\%. TOTAL lies within 3\% of CPU time in 95.5\% of its 111 compiles longer than 2 s, fewer than in S1 and S2; no cell holds more than one of the others, so the cell medians absorb them. Its sentinel stayed within 4.8\% of its median.

\begin{table*}[t]
\small
\caption{The jobs of a session (\S{}3.5).}
\label{tab:jobs}
\begin{tabularx}{\textwidth}{lll>{\raggedright\arraybackslash}X>{\raggedright\arraybackslash}X}
\toprule
Job & Mode & Corpus & Sweep & Arms \\
\midrule
FrontendN & front end & scalar, M = 16 & N $\in$ \{0, 500, 1,000, 2,000, 4,000\} & all light arms, incl. \texttt{manual}, \texttt{xmacro} and \texttt{pfr\_\allowbreak{}native} \\
FrontendHeavy & front end & scalar, M = 16 & N $\in$ \{0, 25, 50, 100, 200\} & reflect-cpp and Glaze, both paths, + \texttt{reflect\_\allowbreak{}plain}, \texttt{p1061\_\allowbreak{}visit}, \texttt{pfr} \\
FrontendM & front end & scalar, N = 50 & M $\in$ \{0, 8, 16, 32, 64, 128\} & light arms except \texttt{manual}, \texttt{xmacro} and \texttt{pfr\_\allowbreak{}native}, + heavy arms \\
RichN & front end & rich, M = 16 & N $\in$ \{0, 25, 50, 100, 200\} & every serializer arm of S1 and S2 \\
HeteroN & front end & hetero & N $\in$ \{0, 200, 400, 800, 1,600\} & light arms except \texttt{xmacro\_\allowbreak{}unique}, which cannot nest \\
HeteroHeavy & front end & hetero & N $\in$ \{0, 25, 50, 100, 200\} & heavy arms + \texttt{reflect\_\allowbreak{}plain}, \texttt{p1061\_\allowbreak{}visit}, \texttt{pfr} \\
FullN & \texttt{-O2} & scalar, M = 16 & N $\in$ \{0, 50, 100, 200, 400\} & light arms except \texttt{xmacro} and \texttt{pfr\_\allowbreak{}native}, + split controls \\
FullHeavy & \texttt{-O2} & scalar, M = 16 & N $\in$ \{0, 25, 50, 100\} & heavy arms + \texttt{reflect\_\allowbreak{}plain}, \texttt{p1061\_\allowbreak{}visit}, \texttt{pfr} \\
OverloadN & front end & scalar, M = 16 & K = C = N $\in$ \{0, 250, 500, 1,000\} & ten idioms \\
OverloadC & front end & scalar, M = 16 & C $\in$ \{0, 250, 500, 1,000, 2,000\} at K = 2,000 (every type also called once) & free function, \texttt{serialize} template, dispatch template, distinct names \\
OverloadK & front end & scalar, M = 16 & K $\in$ \{500, 1,000, 2,000, 4,000\} at C = 500 & free function, \texttt{serialize} template, distinct names \\
CompilerN, CompilerFlag & front end & scalar, M = 16 & N $\in$ \{0, 500, 1,000, 2,000\} & key arms on GCC 16.2.0 and 16.1.0, paired; CompilerFlag with and without \texttt{-freflection} \\
CompilerHeavy & front end & scalar, M = 16 & N $\in$ \{0, 50, 100, 200\} & both paths of Glaze and reflect-cpp + \texttt{reflect\_\allowbreak{}plain}, on both compilers, paired \\
FlagCost, FlagCostHeavy & front end & scalar, M = 16 & N $\in$ \{0, 500, 1,000, 2,000\} / \{0, 50, 100, 200\} & non-reflection arms with and without \texttt{-freflection} \\
FrontendIdiom (S3 only) & front end & scalar, M = 16 & N $\in$ \{0, 500, 1,000, 2,000, 4,000\} & \texttt{reflect\_\allowbreak{}plain}, \texttt{reflect\_\allowbreak{}array}, \texttt{reflect\_\allowbreak{}index}, \texttt{reflect\_\allowbreak{}seq}, \texttt{p1061}, \texttt{pfr}, \texttt{manual\_\allowbreak{}unique} \\
ReportOverhead & front end & scalar, M = 16 & N $\in$ \{0, 1,000, 2,000\} & \texttt{reflect\_\allowbreak{}plain}, \texttt{pfr}, \texttt{manual\_\allowbreak{}unique} with and without \texttt{-ftime-report} (external wall time) \\
\bottomrule
\end{tabularx}
\end{table*}

\subsection{Statistics}\label{sec:3.6}

For each cell the estimate is the median over measured rounds. Per-type costs are ordinary least-squares slopes of the cell medians against N (or M, K, C), for curves that pass the linearity check below. Uncertainty comes from a bootstrap over rounds (4,000 resamples): rounds are drawn with replacement and every statistic --- cell medians, slopes, ratios, shares and crossovers --- is recomputed on the same draw, which keeps the within-round pairing of arms. Because \texttt{-ftime-report} prints wall time in 10 ms steps (GCC 16 can write the same report at finer resolution into SARIF output, which we did not use), each resampled value receives uniform jitter of one step (a smoothed bootstrap), so near-constant cells do not yield zero-width intervals; timer rows GCC omits (under about 5 ms) count as zero. Point estimates use the unsmoothed medians, so an estimate can sit just outside its smoothed interval; this happens for two cited numbers, both by less than 0.01. Intervals are 95\% percentile intervals within one session.

Where a curve is not linear we report point values or a fitted quadratic, not a slope. A curve counts as not linear when the slope over the upper half of its sweep differs from the slope over the lower half by more than a factor of 1.25 and the 95\% interval of that ratio excludes 1, or when its linear fit has R\textsuperscript{2} \textless{} 0.98 and misses some cell by more than three timer steps. The half test is skipped for a curve that rises by less than ten timer steps over either half, where the timer cannot show curvature; the R\textsuperscript{2} test still applies. A curve that passes can still bend, so a slope is an average over its job's sweep and depends on that sweep: \texttt{reflect\_\allowbreak{}plain} costs 1.72 ms per struct over N $\leq$ 4,000 (FrontendN) and 1.39 ms over N $\leq$ 200 (FrontendHeavy), \texttt{pfr} 1.66 and 1.39 ms. This is a further reason why ratios are formed only within one job.

Unless marked otherwise, every measured number in \S{}4 comes from session S1; the native formulations of \S{}4.2 come from S3 (job FrontendIdiom), and the Boost.PFR probe of \S{}4.3 is a separate run. Session S2 repeats the full plan on another day with the same compilers, code and protocol. Two sessions cannot estimate between-session variance, so we do not pool intervals across sessions; \S{}4.6 instead compares the sessions arm by arm, as the ratio of each per-type slope in S2 to its value in S1, and gives the pooled point estimates of the main ratios. The number of rounds is checked afterwards by resampling fewer rounds: with five rounds per cell instead of eleven, the 95\% interval of the FrontendN slope widens from 1.5\% to 2.1\% of the estimate for \texttt{reflect\_\allowbreak{}plain}, from 2.2\% to 3.3\% for \texttt{p1061} and from 1.4\% to 2.0\% for \texttt{pfr}.

\subsection{Toolchain, libraries, machine}\label{sec:3.7}

GCC 16.2.0 built from the release tarball (SHA-512 checked against the published sum, signature checked) with the configure options of the GCC 16.1.0 used for comparison, both non-bootstrapped and built by the same host compiler (MSYS2 GCC 15.2.0), with the same install layout (the MinGW headers in the prefix, the compiler executables stripped of debug information), \texttt{-{}-enable-checking=\allowbreak{}release}, flags \texttt{-std=\allowbreak{}c++26 -freflection}. Both compilers are configured with \texttt{-{}-with-tune=\allowbreak{}native}, which the driver resolves to \texttt{-mtune=\allowbreak{}alderlake} on this machine. Libraries were pinned by commit to their latest releases at the time (2026-11-17) and fetched as sha256-verified archives: Boost.PFR from Boost 1.92.0 (401385c)~\cite{boostpfr}, reflect-cpp v0.27.1 (e0ee3f6)~\cite{reflectcpp}, Glaze v9.1.0 (43b175e)~\cite{glaze}. Machine: Intel Core i5-14600KF (6 performance + 8 efficiency cores; the harness and the compiler run on the 12 logical CPUs of the performance cores), 32 GB, Windows 11 (build 26300) with Microsoft Defender real-time protection on, MinGW-w64 UCRT.

\section{Results}\label{sec:4}

\subsection{Fixed costs (RQ4)}\label{sec:4.1}

Including \texttt{\textless{}meta\textgreater{}} and the native arm's headers adds 30 ms [23, 37] to a translation unit that already includes \texttt{\textless{}string\textgreater{}}, \texttt{\textless{}vector\textgreater{}} and \texttt{\textless{}optional\textgreater{}} (\texttt{reflect\_\allowbreak{}plain} against \texttt{p1061} at N = 0, same job). The libraries' headers (the narrowest that provide each visitor, \S{}3.1) cost far more. Over Boost.PFR at N = 0, reflect-cpp adds 620 ms on its C++20 path and 550 ms on its C++26 path; Glaze adds 1260 ms with its traditional backend and 1270 ms with its P2996 backend. These fixed costs are excluded from every per-type number below by construction. Whether added headers also change per-type cost is what the code-identical control measures (\S{}4.2).

\textbf{The \texttt{-freflection} flag.} On code that does not use reflection, enabling the flag raises per-type front-end cost by a factor of 1.188 for hand-written code, 1.239 for P1061 and 1.082 for Boost.PFR, and by 1.039 and 1.017 for the pre-C++26 paths of Glaze and reflect-cpp. It raises fixed cost by one to three 10 ms timer steps (Boost.PFR 10 ms [1, 16], Glaze 30 ms [22, 51]). The compiled code is identical with and without it (\S{}3.3). On GCC 16.1.0 the flag costs far more (\S{}4.6). GCC 16.0 had a 19\% flag cost on \texttt{\textless{}bits/\allowbreak{}stdc++.\allowbreak{}h\textgreater{}} (PR124925~\cite{gccpr124925}), and GCC 16.2.0 contains, among other reflection changes, a rewrite of the check responsible for a 40-fold slowdown in another case (PR125179~\cite{gccpr125179}). Every serializer arm in this study is compiled with the flag (only the \texttt{*\_\allowbreak{}noflag} controls are not), so the native arms carry no flag-only handicap in the comparisons below; a project that turns the flag on to use reflection also pays its cost on the code that does not use it.

\textbf{The instrument.} Compiling with \texttt{-ftime-report} raises the per-struct slope of external wall time by 1.028$\times$ for the native arm, 1.041$\times$ for Boost.PFR and 1.123$\times$ [1.094, 1.150] for hand-written code (whole translation units at N = 2,000: 1.031, 1.042 and 1.086). Relative to its cost, hand-written code pays the most, so ratios against it read lower with the report: on external wall time the native arm costs 6.92$\times$ hand-written code per struct with the report and 7.56$\times$ without it, while its ratio to Boost.PFR is the same either way (1.00 and 1.01).

\subsection{Per-type cost of the mechanisms (RQ1)}\label{sec:4.2}

\begin{table}[t]
\small
\caption{Front-end cost per struct (scalar corpus, M = 16, FrontendN, N up to 4,000; session S1).}
\label{tab:1}
\begin{tabularx}{\columnwidth}{>{\raggedright\arraybackslash}Xrl}
\toprule
Arm & ms per struct & 95\% CI \\
\midrule
\texttt{manual\_\allowbreak{}unique} & 0.23 & [0.23, 0.23] \\
\texttt{xmacro\_\allowbreak{}unique} & 0.24 & [0.24, 0.25] \\
\texttt{p1061} & 0.22 & [0.21, 0.22] \\
\texttt{p1061\_\allowbreak{}visit} & 0.45 & [0.45, 0.46] \\
\texttt{reflect\_\allowbreak{}plain} & 1.72 & [1.70, 1.73] \\
\texttt{reflect\_\allowbreak{}visit} & 2.01 & [2.00, 2.02] \\
\texttt{reflect} & 2.36 & [2.32, 2.37] \\
\texttt{pfr} & 1.66 & [1.65, 1.67] \\
\texttt{pfr\_\allowbreak{}native} (code-identical to \texttt{pfr}) & 1.77 & [1.75, 1.78] \\
\bottomrule
\end{tabularx}
\end{table}

Boost.PFR's C++17 path (\texttt{pfr\_\allowbreak{}classic}) has no row: its cost is not linear in N (\S{}4.3). Neither are \texttt{manual} and \texttt{xmacro}, which share one overloaded name (\S{}4.4). The other rows pass the linearity check of \S{}3.6. Still, the per-struct cost of the P2996 arms (all in the span form; the other forms of S3 do not rise, below), of \texttt{p1061\_\allowbreak{}visit} and of Boost.PFR rises with N, so their slopes are averages over the sweep. Inline P1061 and hand-written code stay flat: 0.20 and 0.21 ms per struct over the first 500 and the last 2,000 structs, and 0.22 and 0.23 ms. The native arm's marginal cost per struct, (cost at N $-$ cost at N = 0) / N, is 1.44 ms at N = 500 and 1.71 ms at N = 4,000. For Boost.PFR it is 1.44 and 1.65 ms. The rise is not specific to P2996: the native arm's upper-half slope is 1.14$\times$ [1.12, 1.17] its lower-half slope, and that of \texttt{p1061\_\allowbreak{}visit} 1.23$\times$. In the span form, the concurrent name-lookup timer of a whole translation unit grows from 0.12 s at N = 1,000 to 0.98 s at 4,000. That is about as much as the native arm's rise above a constant per-struct cost. GCC 16.2.0 declares the range, begin and per-element iterator variables of a constant iterating expansion statement \texttt{static} (2 + n of them per expansion over n elements). It registers every block-scope \texttt{static} by a linear scan of one list for the whole translation unit (\texttt{determine\_\allowbreak{}local\_\allowbreak{}discriminator} in \texttt{cp/\allowbreak{}decl.\allowbreak{}cc}). Each expansion therefore pays for all those compiled before it. Ordinary block-scope statics show the same quadratic growth without any reflection, on every GCC release from 11.5 to the current development branch. This is a GCC compile-time issue, not a property of P2996, reported as PR128928~\cite{gccpr128928} (reproducers in \texttt{Repro/\allowbreak{}GccLocalStatics}). In S2 the slopes of Table~\ref{tab:1} are 1.000$\times$ (\texttt{manual\_\allowbreak{}unique}), 1.009$\times$ (\texttt{p1061}), 1.002$\times$ (\texttt{pfr}), 0.990$\times$ (\texttt{reflect\_\allowbreak{}plain}) and 0.995$\times$ (\texttt{reflect}) their S1 values.

\begin{figure*}[t]
\centering
\includegraphics[width=\textwidth]{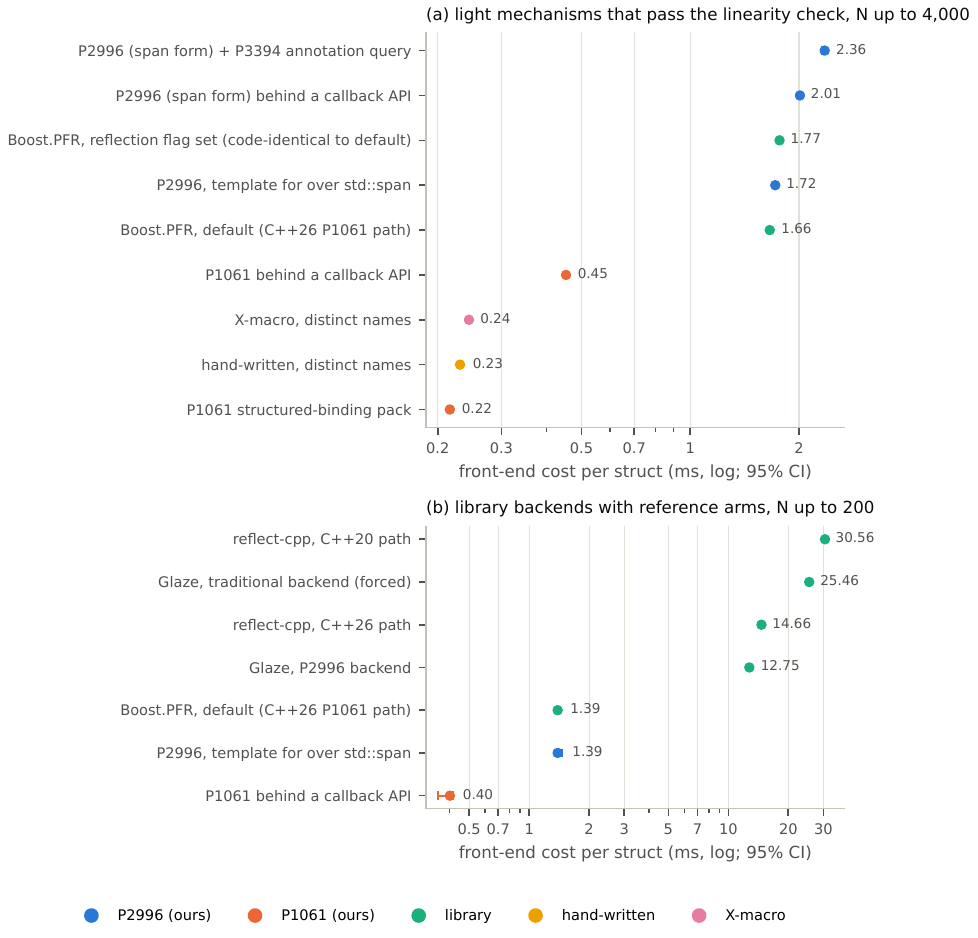}
\Description{Front-end cost per struct, scalar corpus, M = 16, session S1, log scale}
\caption{Front-end cost per struct, scalar corpus, M = 16, session S1, log scale. The P2996 arms use the span form of Listing~\ref{lis:1}. Each dot is the least-squares slope of the cell medians (eleven measured rounds) against N; whiskers are 95\% bootstrap intervals and are mostly narrower than the dot. (a) Job FrontendN, N = 0--4,000: the arms of Table~\ref{tab:1}, all of which pass the linearity check of \S{}3.6; \texttt{manual}, \texttt{xmacro} and Boost.PFR's C++17 path are not linear in N and are omitted. (b) Job FrontendHeavy, N = 0--200: both paths of reflect-cpp and Glaze, with the native arm, Boost.PFR and \texttt{p1061\_\allowbreak{}visit} as references. The reference arms cost less per struct in (b) than in (a) because their per-struct cost rises with the number of types already in the translation unit; ratios are therefore taken within one panel only.}
\label{fig:1}
\end{figure*}

\textbf{Mechanism $\times$ API shape.} Here, as in every job but FrontendIdiom, P2996 enumeration means the span form of Listing~\ref{lis:1}, and its formulation alone moves its cost by a factor of about two (S3, below). With the serializer written inline, P2996 enumeration costs 7.96$\times$ [7.83, 8.04] as much per struct as a P1061 pack (7.20$\times$ over the first 500 structs, 8.47$\times$ over the last 2,000). Behind a callback API the ratio is 4.44$\times$ [4.40, 4.48]. The callback shape itself costs 1.17$\times$ with P2996 and 2.10$\times$ with P1061. Boost.PFR builds on the same P1061 primitive behind the same callback shape, and its own machinery puts it at 3.66$\times$ [3.63, 3.70] of \texttt{p1061\_\allowbreak{}visit}: under \texttt{-std=\allowbreak{}c++26} its \texttt{for\_\allowbreak{}each\_\allowbreak{}field} binds the pack, builds an index sequence of the field count, binds the pack again and calls, for every field, a dispatcher (\texttt{for\_\allowbreak{}each\_\allowbreak{}field\_\allowbreak{}impl\_\allowbreak{}apply}) that receives the field index as a type (\texttt{size\_\allowbreak{}t\_\allowbreak{}\textless{}I\textgreater{}}) and first tests by SFINAE whether the callback also accepts that index.

\textbf{Native enumeration against Boost.PFR.} Written inline in the span form, native enumeration costs about as much per struct as Boost.PFR's \texttt{for\_\allowbreak{}each\_\allowbreak{}field}. This parity describes the choice between an inline native serializer in the span form and Boost.PFR's visitor, not the two enumeration mechanisms. The slope ratio is 1.04$\times$ [1.02, 1.04]. The ratio of marginal costs depends on N, and so does which arm is cheaper: 0.95$\times$ [0.94, 0.97] at N = 1,000, 0.99$\times$ [0.98, 1.01] at 2,000 and 1.03$\times$ [1.02, 1.04] at 4,000. The slope ratio is 0.99$\times$ in the CompilerN job (N $\leq$ 2,000), 0.96$\times$ on the rich corpus and 0.99$\times$ on the heterogeneous one (1.07$\times$ [1.05, 1.09] over N $\leq$ 200, HeteroHeavy; \S{}4.6). All of these differences are about as large as the code-identical pair's below (1.03--1.06$\times$) or smaller. With the API shape matched, native enumeration behind the same callback (\texttt{reflect\_\allowbreak{}visit}) costs 1.21$\times$ [1.20, 1.22] as much as Boost.PFR per struct (S2: 1.21), 1.14$\times$ on the rich corpus and 1.08$\times$ on the heterogeneous graph. On every corpus this exceeds the code-identical pair's difference, although the native arm also parses \texttt{\textless{}meta\textgreater{}}, which that pair suggests is worth a few percent. The 1.21 is the quotient of the two decompositions above. Behind a callback, P2996 enumeration costs 4.44$\times$ P1061's, while Boost.PFR's machinery costs 3.66$\times$ the same primitive. Native enumeration matches Boost.PFR by being written inline, which avoids the callback shape's factor of 1.17. Written inline over a constant array instead of the span, it costs about half as much as Boost.PFR's \texttt{for\_\allowbreak{}each\_\allowbreak{}field} (below), again without matching the API shape.

\begin{figure*}[t]
\centering
\includegraphics[width=\textwidth]{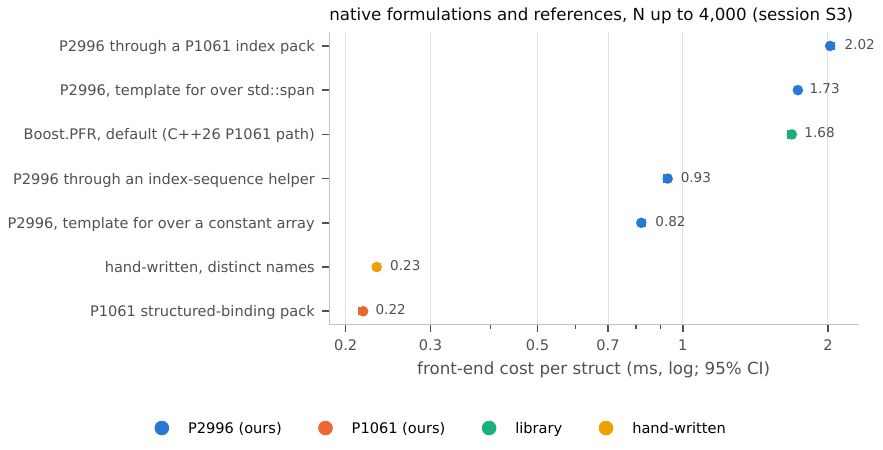}
\Description{Front-end cost per struct of four native formulations of the same member query and three reference arms, scalar corpus, M = 16, session S3 (job FrontendIdiom, N = 0--4,000), log scale}
\caption{Front-end cost per struct of four native formulations of the same member query and three reference arms, scalar corpus, M = 16, session S3 (job FrontendIdiom, N = 0--4,000), log scale. Each dot is the least-squares slope of the cell medians (eleven measured rounds) against N; whiskers are 95\% bootstrap intervals. The span form is the native arm of every other job.}
\label{fig:2}
\end{figure*}

\textbf{Native formulation (S3).} The native arms above enumerate with one form, used in several of P2996R13's own examples~\cite{wg21p2996} (e.g. its universal formatter): an expansion statement over the \texttt{std::\allowbreak{}span} that \texttt{define\_\allowbreak{}static\_\allowbreak{}array} returns. C++26 makes this an iterating expansion statement, defined element by element through the span's iterator, and GCC 16.2.0 evaluates it that way. The compiler constant-evaluates a loop over the range to count its elements, then for each field resolves \texttt{begin + i} and its dereference on \texttt{std::\allowbreak{}span}'s class-type iterator, deduces \texttt{auto} and constant-initializes a \texttt{static constexpr} iterator and the member. Over an array, C++26 expands by destructuring. Session S3 times the same member query in three other forms, in one job with the span form (FrontendIdiom, N $\leq$ 4,000; Figure~\ref{fig:2}). Per struct, the same expansion statement over a constant array (\texttt{[:reflect\_\allowbreak{}constant\_\allowbreak{}array(\allowbreak{}...):]}) costs 0.82 against 1.73 ms over the span, 0.47$\times$ [0.47, 0.48] (as marginal costs, 0.53$\times$ at N = 1,000 and 0.48$\times$ at 4,000, since only the span form bends). With fixed costs included, a whole translation unit with the array form takes 0.61$\times$ the span form's front-end time at N = 1,000 and 0.50$\times$ at 4,000. An index-sequence helper costs 0.93 ms (0.54$\times$), and reflect-cpp's form, a \texttt{constexpr} P1061 index pack bound through the tuple protocol that C++26 gives \texttt{std::\allowbreak{}integer\_\allowbreak{}sequence} (P1789R3), 2.02 ms (1.17$\times$). That form's overload-resolution timer covers 54\% of its time, consistent with one \texttt{get\textless{}j\textgreater{}} resolved by argument-dependent lookup per element.

About half of the span form's per-struct cost therefore comes with the span, not the member query. The contrast also removes the calls that \texttt{define\_\allowbreak{}static\_\allowbreak{}array} makes around \texttt{reflect\_\allowbreak{}constant\_\allowbreak{}array} (\texttt{type\_\allowbreak{}of}, \texttt{is\_\allowbreak{}array\_\allowbreak{}type}, \texttt{extract}, \texttt{extent} and the \texttt{span} constructor), which we did not time separately. They run once per struct, whereas the iterating expansion's per-element steps run once per field, so we attribute the difference mainly to the iterating expansion. The expansion statement as such is not shown to be costly: the two forms without it differ in direction, and the cheapest of the four forms is itself an expansion statement, the one over the array (0.88$\times$ the helper). With the array form, inline native enumeration costs 3.77$\times$ inline P1061 and 0.49$\times$ Boost.PFR's callback \texttt{for\_\allowbreak{}each\_\allowbreak{}field} per struct (S3 has no callback-shaped array arm). It adds 0.59 ms over hand-written code and does not rise with N (upper-half / lower-half slope 0.99, against 1.15 over the span), nor do the other two forms (0.97 for the P1061 index pack, 0.98 for the index-sequence helper). The arms that also run in FrontendN reproduce their S1 values: 1.73 against 1.72 ms for the span form, 1.68 against 1.66 ms for Boost.PFR, 0.22 against 0.22 ms for P1061, 0.23 against 0.23 ms for hand-written code, and 7.97$\times$ against 7.96$\times$ for native / P1061. S3 is a single session, so its intervals understate day-to-day variation, which for the slopes timed in both S1 and S2 had a median of 0.7\% and at most 8.8\% (\S{}4.6), far below the factor of two between the span and array forms. S3 timed the formulations at the front end only, on the scalar corpus at M = 16; every other native number in this paper uses the span form.

\textbf{The annotation query.} Querying \texttt{annotations\_\allowbreak{}of\_\allowbreak{}with\_\allowbreak{}type} on every member makes the native serializer 1.37$\times$ [1.36, 1.38] as costly per struct of 16 scalar fields. In this corpus no member is annotated, so the query is pure overhead that buys the ability to skip annotated fields.

\textbf{Where the time goes.} GCC 16 has three kinds of timer. Phases (\texttt{phase parsing}, \texttt{phase lang.\allowbreak{} deferred}, \texttt{phase opt and generate}) partition the run. \texttt{template instantiation} and \texttt{constant expression evaluation} are stack timers that exclude each other. \texttt{\textbar{}name lookup} and \texttt{\textbar{}overload resolution} are concurrent timers: they keep running through the deduction, substitution and instantiation that a lookup or a call triggers, and so overlap all others. Constant evaluation accounts for 43\% of the native arm's per-struct time and template instantiation for 46\%, together most of it. For Boost.PFR the shares are 1\% and 84\%, and for P1061 1\% and 79\%. Both native shares include the expansion statement's per-element steps, not only the reflection queries. In S3 the span form spends 46\% of its 1.73 ms in template instantiation and 43\% in constant evaluation; the array form spends 35\% and 53\% of its 0.82 ms. More than half of the difference between the two forms therefore lies in template instantiation. We did not time the P2996 operations separately.

\textbf{Memory} ranks the arms differently from time. Per struct, the compiler allocates 266 KB of GGC memory for the native arm and its peak working set grows by 235 KB; for Boost.PFR, at about the same time per struct, 426 and 408 KB; for P1061 60 and 50 KB; and for hand-written code 73 and 60 KB. The annotation query raises the native arm's to 382 and 333 KB. GGC memory is identical in every round of a cell, and the medians of the peak working set nearly so: the per-struct intervals of both are narrower than 0.1 KB.

\textbf{Hand-written code} with a distinct name per type is one of the three cheapest arms, with the P1061 pack and the X-macro with distinct names (Table~\ref{tab:1}). Per struct, P1061 costs 0.94$\times$ [0.92, 0.94] and the X-macro 1.06$\times$ [1.04, 1.06] as much as the hand-written arm. We did not measure the instrument's overhead for these two arms; on the hand-written arm it raises the per-struct wall time by 1.123$\times$ (\S{}4.1), more than these differences, so we read the three arms as level rather than ranked. On the scalar corpus, native enumeration costs 7.46$\times$ as much per struct as hand-written code, a ratio that would be about 9\% larger without \texttt{-ftime-report} (\S{}4.1), and 10.22$\times$ with the annotation query, for which we did not measure the instrument's overhead.

\textbf{The code-identical control.} \texttt{pfr\_\allowbreak{}native} compiles to the same code as \texttt{pfr} (\S{}3.3) and allocates the same GGC memory per struct (426 KB); its flag only switches Boost.PFR to reflection-based definitions that \texttt{for\_\allowbreak{}each\_\allowbreak{}field} never instantiates, in headers that include \texttt{\textless{}meta\textgreater{}}. Yet its per-struct time is 1.06$\times$ [1.05, 1.07] that of \texttt{pfr}, over the first 500 structs (1.04$\times$) as over the last 2,000 (1.07$\times$), and the same pooled over both sessions (1.06$\times$). On the heterogeneous corpus it is 1.03$\times$ [1.01, 1.04]; on the rich corpus, at N $\leq$ 200, it is not resolved (0.97$\times$ [0.95, 1.02]). So what a translation unit parses can change the per-type cost of unchanged code by a few percent, not only its fixed cost. The native arms parse \texttt{\textless{}meta\textgreater{}}; the P1061, hand-written and Boost.PFR arms do not. Differences of a few percent between them therefore cannot be attributed to the enumeration mechanism. We did not test whether \texttt{\textless{}meta\textgreater{}} alone causes the effect.

\subsection{Library backends (RQ2)}\label{sec:4.3}

\begin{table}[t]
\small
\caption{Same-source backend A/Bs: per-struct front-end cost in ms, previous path / C++26 path (scalar corpus, M = 16). reflect-cpp and Glaze rows: slopes over N = 0--200 (FrontendHeavy); Boost.PFR rows: marginal cost per struct at the stated N (FrontendN)\textsuperscript{a}.}
\label{tab:2}
\begin{tabularx}{\columnwidth}{>{\raggedright\arraybackslash}Xrrrl}
\toprule
Library & previous path & C++26 path & ratio & 95\% CI \\
\midrule
reflect-cpp & 30.56 & 14.66 & 2.08 & [2.06, 2.10] \\
Glaze & 25.46 & 12.75 & 2.00 & [1.96, 2.01] \\
Boost.PFR, N = 500 & 6.32 & 1.44 & 4.4 & [4.3, 4.5] \\
Boost.PFR, N = 4,000 & 21.07 & 1.65 & 12.8 & [12.7, 12.9] \\
\bottomrule
\end{tabularx}
\par\smallskip\footnotesize \textsuperscript{a} Boost.PFR's C++17 path is not linear in N, so it has no per-struct slope. Its rows give the marginal cost per struct at the stated N, (cost at N $-$ cost at N = 0) / N, for both paths, and their ratio.
\end{table}

\begin{figure*}[t]
\centering
\includegraphics[width=\textwidth]{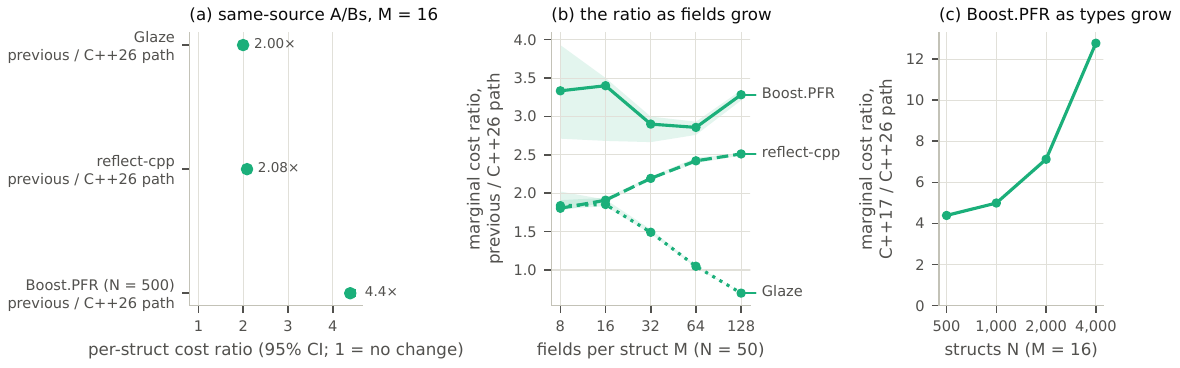}
\Description{Previous path / C++26 path for each library, scalar corpus, session S1}
\caption{Previous path / C++26 path for each library, scalar corpus, session S1. (a) M = 16: for reflect-cpp and Glaze the ratio of per-struct slopes over N = 0--200 (FrontendHeavy, Table~\ref{tab:2}), for Boost.PFR the ratio of marginal costs at N = 500 (FrontendN); whiskers are 95\% bootstrap intervals and the vertical line marks a ratio of 1. (b) Field sweep at N = 50 (FrontendM): the ratio of the cost the fields add, cost at M minus cost at M = 0, with 95\% bands; Boost.PFR's C++26 path adds fewer than ten timer steps at M = 8 and 16, hence its wide band there. Because the field-independent cost is excluded and N is fixed at 50, the values at M = 16 differ from (a). (c) Boost.PFR's ratio of marginal costs against N (FrontendN, M = 16).}
\label{fig:3}
\end{figure*}

\textbf{Boost.PFR's C++17 path is quadratic in the number of types.} From N = 2,000 to 4,000 its marginal front-end cost multiplies by 3.7 (C++26 path: 2.0), reaching 84 s at N = 4,000, of which the overload-resolution timer accounts for 87\%; its GGC memory grows linearly (1262 KB per struct), and the whole translation unit peaks at 5.0 GB of working set at N = 4,000 (C++26 path: 1.6 GB). On the rich corpus, whose sweep ends at N = 200, the curve is still close to a line ($\times$2.2 from N = 100 to 200); on the heterogeneous corpus it already bends within the sweep ($\times$2.6 from N = 800 to 1,600; upper-half slope 1.59$\times$ [1.55, 1.64] the lower-half slope, not linear by \S{}3.6). The quadratic term therefore matters for translation units that reflect many types. A separate probe (\texttt{Harness/\allowbreak{}pfrprobe.\allowbreak{}py}; compiles one at a time on the performance cores, medians of three runs) locates it.

\begin{itemize}
\item \emph{Not GCC 16 and not the flag.} The path grows alike on GCC 15.2.0 with \texttt{-std=\allowbreak{}c++20} (7, 21 and 78 s at N = 1,000, 2,000 and 4,000), and on GCC 16.2.0 with \texttt{-std=\allowbreak{}c++20} and with \texttt{-std=\allowbreak{}c++26}, with and without \texttt{-freflection}. The overload-resolution timer, which also runs through the deduction and substitution that resolution triggers, covers 87\% to 89\% of the time at N = 4,000 in every case.
\item \emph{Field counting.} On GCC 16.2.0 with \texttt{-std=\allowbreak{}c++20}, each test calls one piece of the library for every struct. Field counting alone (\texttt{detail::\allowbreak{}fields\_\allowbreak{}count}) reproduces the growth: its overload-resolution time is 1.47, 4.42 and 15.83 s at N = 500, 1,000 and 2,000, against 18.01 s for the whole \texttt{for\_\allowbreak{}each\_\allowbreak{}field} at N = 2,000. Binding the fields with the count given (\texttt{detail::\allowbreak{}tie\_\allowbreak{}as\_\allowbreak{}tuple}) takes 0.91 s at N = 2,000.
\end{itemize}

The library counts the 16 fields with about ten aggregate-initialization probes per type (\S{}2). Each probe is a function template whose return type is \texttt{std::\allowbreak{}add\_\allowbreak{}pointer\_\allowbreak{}t\textless{}decltype(\allowbreak{}T\{ubiq\{I\}...\})\textgreater{}}, guarded by SFINAE and overloaded with an rvalue variant. One such probe per type already grows quadratically: 0.30, 1.05 and 4.21 s at N = 500, 1,000 and 2,000; a copy of it takes 4.15 s at N = 2,000. Dropping the \texttt{std::\allowbreak{}add\_\allowbreak{}pointer\_\allowbreak{}t} wrapper from the copy cuts that to 0.97 s, and dropping the rvalue overload and the guard as well leaves 0.33 s; both reduced probes still grow faster than linearly from N = 1,000 to 2,000. We did not identify the cause inside GCC, and we did not test Clang.

\textbf{As fields grow.} The ratio is not a constant of the library. In the field sweep (N = 50), Glaze's traditional / P2996 ratio is 1.84 at M = 8, 1.85 at 16, 1.49 at 32, 1.05 at 64 and 0.70 at 128 (S2: 0.70); reflect-cpp's is 1.91 at 16 and 2.51 at 128; Boost.PFR's lies between 2.86 (M = 64) and 3.28 (M = 128) from M = 32 on, with 2.90 at 32. From M = 64 to 128, the cost the fields add grows 3.46$\times$ for Glaze's P2996 backend against 2.31$\times$ for its traditional backend. At M = 128 the P2996 backend's whole translation unit (50 structs) peaks at 5.0 GB of working set against 3.0 GB, so memory reverses with time. A mechanism consistent with Glaze's trend (a hypothesis we did not isolate): at v9.1.0 its P2996 backend calls \texttt{nonstatic\_\allowbreak{}data\_\allowbreak{}members\_\allowbreak{}of} once per member index, both for access (\texttt{member\_\allowbreak{}info\_\allowbreak{}at\textless{}T,\allowbreak{} I\textgreater{}}) and for names (\texttt{get\_\allowbreak{}member\_\allowbreak{}name\_\allowbreak{}p2996\textless{}T,\allowbreak{} I\textgreater{}}), so its per-struct reflection work grows faster than linearly in M. Glaze's main branch has since changed this code (PR \#3112~\cite{glazepr3112}, after v9.1.0), so the trend is a property of the pinned release. The traditional backend is superlinear in M as well, because its member count probes aggregate initialization with 1, 2, \ldots{}, M + 1 initializers (\S{}2). The hypothesis therefore concerns why the P2996 backend's curve bends more. reflect-cpp's C++20 path reaches 8.4 GB at M = 128 (C++26 path: 3.1 GB). At M = 16 the field-sweep ratios lie below Table~\ref{tab:2}'s, which are ratios of slopes over N $\leq$ 200. The backend ratio also depends on N: in FrontendHeavy the ratio of marginal costs is 1.86 at N = 50 and 1.98 at N = 200 for Glaze, and 1.81 and 2.05 for reflect-cpp.

\textbf{Libraries against native enumeration.} In the same job (FrontendHeavy, N $\leq$ 200, where the native arm costs 1.39 ms per struct against Table~\ref{tab:1}'s 1.72 ms over N $\leq$ 4,000), reflect-cpp's C++26 path costs 10.55$\times$ and Glaze's P2996 backend 9.17$\times$ the native arm per struct. Both libraries compute field names on the path we call although our serializer ignores them --- Glaze's \texttt{for\_\allowbreak{}each\_\allowbreak{}field} reads \texttt{reflect\textless{}T\textgreater{}::\allowbreak{}size}, which is the size of \texttt{member\_\allowbreak{}names\textless{}T\textgreater{}}, and reflect-cpp's \texttt{to\_\allowbreak{}view} builds a named tuple --- so these ratios measure each library's per-type work for its field visitor, not an intrinsic cost of library-based reflection. Per struct, the C++26 paths of Glaze and reflect-cpp also roughly halve the growth of the compiler's peak working set: 2,494 against 4,287 KB for Glaze and 2,846 against 6,228 KB for reflect-cpp.

\textbf{Whole translation units.} Per-type ratios exclude fixed costs. In a whole translation unit, switching to the C++26 path saves less, because the headers of Glaze and reflect-cpp are paid once: 1.26$\times$ (Glaze) and 1.47$\times$ (reflect-cpp) of the front-end time with 50 types, and 1.61$\times$ and 1.84$\times$ with 200 types (Table~\ref{tab:3}).

\begin{table*}[t]
\small
\caption{Whole translation unit at the front end (scalar corpus, M = 16, FrontendHeavy, session S1): the compiler's TOTAL and its peak working set, without structs and with 200.}
\label{tab:3}
\begin{tabularx}{\textwidth}{>{\raggedright\arraybackslash}Xrrrr}
\toprule
Arm & TOTAL, N = 0 (s) & TOTAL, N = 200 (s) & peak, N = 0 (MB) & peak, N = 200 (MB) \\
\midrule
\texttt{reflect\_\allowbreak{}plain} & 0.31 & 0.59 & 93 & 140 \\
\texttt{pfr} & 0.28 & 0.56 & 86 & 163 \\
\texttt{rflcpp} & 0.90 & 7.10 & 252 & 1480 \\
\texttt{rflcpp\_\allowbreak{}native} & 0.83 & 3.86 & 209 & 783 \\
\texttt{glaze\_\allowbreak{}classic} & 1.54 & 6.65 & 473 & 1301 \\
\texttt{glaze\_\allowbreak{}native} & 1.55 & 4.13 & 483 & 962 \\
\bottomrule
\end{tabularx}
\end{table*}

\subsection{Per-type serializer idioms and overload sets (RQ3)}\label{sec:4.4}

With one overloaded name per type, hand-written serializers stop being linear: from N = 2,000 to 4,000 the marginal cost of \texttt{manual} multiplies by 4.7 (distinct names: 2.0), reaching 26.6 s at N = 4,000, of which the compiler's overload-resolution timer accounts for 94\%. Its marginal cost is 0.68$\times$ native reflection's at N = 1,000 and 1.33$\times$ at N = 2,000 (against P2996 without the annotation query: 0.92$\times$ and 1.79$\times$), so it overtakes native reflection in the span form between N = 1,000 and 2,000. We do not narrow the crossing further: the measured growth exceeds a quadratic, so a quadratic model would misplace it. The X-macro with one overloaded name behaves alike ($\times$4.7 per doubling).

\begin{table*}[t]
\small
\caption{Ten customization idioms, K = C = N (OverloadN): marginal front-end time at N = 1,000, its growth from N = 500 to 1,000 (2 = linear, 4 = quadratic), and its ratio to distinct names at N = 1,000 (for the idioms that share one lookup set, that ratio itself grows with N).}
\label{tab:4}
\begin{tabularx}{\textwidth}{>{\raggedright\arraybackslash}X>{\raggedright\arraybackslash}Xrrr}
\toprule
Idiom & Form documented by & s at N = 1,000 & $\times$ per doubling & $\div$ distinct names at N = 1,000 \\
\midrule
free function, qualified call & --- & 1.42 & 3.5 & 5.92 \\
free function, ADL, called directly & declared like nlohmann \texttt{to\_\allowbreak{}json}, Boost.JSON \texttt{tag\_\allowbreak{}invoke}, fmt \texttt{format\_\allowbreak{}as} & 1.41 & 3.5 & 5.88 \\
free function, ADL from a library dispatch template & the call shape of nlohmann, Boost.JSON, fmt & 1.48 & 3.5 & 6.17 \\
function template guarded by \texttt{enable\_\allowbreak{}if} (SFINAE), ADL & nlohmann v3.12 \texttt{NON\_\allowbreak{}INTRUSIVE} & 4.97 & 4.0 & 20.71 \\
\texttt{serialize(\allowbreak{}Archive\&,\allowbreak{} T\&,\allowbreak{} unsigned)} template & Boost.Serialization, cereal external & 1.80 & 3.6 & 7.50 \\
one namespace per type, ADL & --- & 0.25 & 2.1 & 1.04 \\
hidden friend (non-template) & \texttt{BOOST\_\allowbreak{}DESCRIBE\_\allowbreak{}CLASS}; nlohmann v3.12 \texttt{INTRUSIVE} declares an \texttt{enable\_\allowbreak{}if}-guarded friend template & 0.29 & 2.2 & 1.21 \\
member function & cereal internal, protobuf & 0.26 & 2.0 & 1.08 \\
class-template specialization & \texttt{glz::\allowbreak{}meta}, \texttt{fmt::\allowbreak{}formatter} & 0.26 & 2.2 & 1.08 \\
distinct names & --- & 0.24 & 2.0 & 1 \\
\bottomrule
\end{tabularx}
\end{table*}

Of the idioms that avoid a shared lookup set, only the hidden friend has a growth per doubling whose interval excludes 2 (2.2 [2.1, 2.3]); at N = 1,000 it costs 1.21$\times$ [1.16, 1.27] as much as distinct names, far below the idioms that share one set.

The costliest idiom is the function template guarded by \texttt{enable\_\allowbreak{}if}, the declaration form of nlohmann::json v3.12's \texttt{NON\_\allowbreak{}INTRUSIVE} macro. It costs far more than the \texttt{serialize} template, which is a function template as well but has no guard (Table~\ref{tab:4}). At N = 1,000 the overload-resolution timer, which also runs through the template argument deduction and substitution that resolution triggers, covers 93\% of its marginal time, template instantiation 26\% and constant evaluation 10\% (plain free functions: 82\%, 1\% and 0\%), and its GGC memory grows by 815 MB against 135 MB for plain free functions. We did not isolate which step of resolving a guarded candidate costs the most.

\begin{figure*}[t]
\centering
\includegraphics[width=\textwidth]{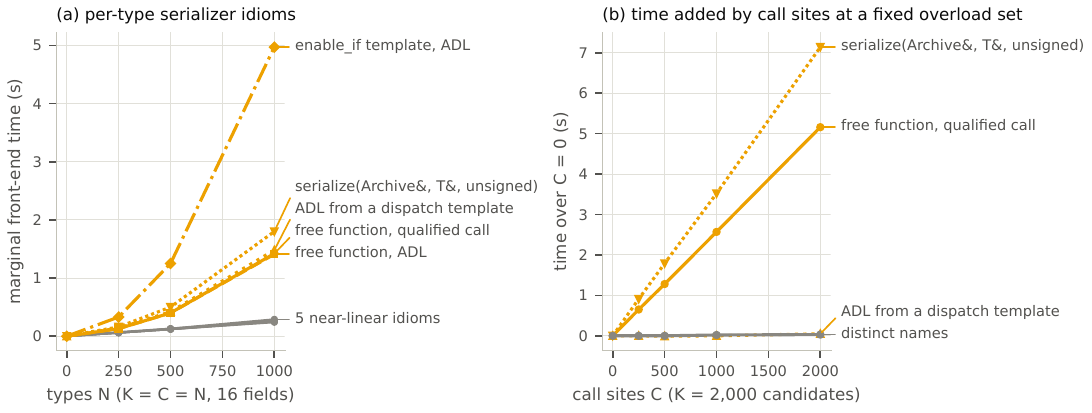}
\Description{Front-end time of the per-type serializer idioms, scalar corpus, M = 16, session S1; medians over eleven measured rounds, no intervals drawn}
\caption{Front-end time of the per-type serializer idioms, scalar corpus, M = 16, session S1; medians over eleven measured rounds, no intervals drawn. (a) OverloadN, K = C = N: marginal time, cost at N minus cost at N = 0, for the ten idioms of Table~\ref{tab:4}. The five idioms whose marginal time at least triples from N = 500 to 1,000 are drawn in color; the qualified free-function call, the direct ADL call and ADL from a dispatch template lie within one 10 ms timer step of each other at N = 250 and within 5\% at N = 500 and 1,000, and overlap. The other five idioms are gray. (b) OverloadC, K = 2,000 overloads, every type already called once: the time added by call sites, cost at C minus cost at C = 0.}
\label{fig:4}
\end{figure*}

\textbf{Call sites and candidates.} At a fixed set of K = 2,000 overloads, with every type already called once outside the swept calls (so serializer bodies are instantiated regardless of C), each further call site costs 2.58 ms with free functions and 3.56 ms with \texttt{serialize} templates, against 0.02 ms with distinct names. At a fixed C = 500 call sites the cost per further overload rises with K: from 0.84 ms (K = 500--1,000) to 1.46 ms (K = 2,000--4,000) with free functions and from 0.96 to 1.67 ms with \texttt{serialize} templates, against 0.22 and 0.23 ms with distinct names. The cost therefore grows in proportion to call sites and faster than linearly with the number of candidates in one lookup set; it is not a property of hand-written code or of the number of types alone. A library that makes the unqualified call inside its own dispatch template resolves it once per type, when the template is instantiated, so it pays per type, not per call site: with every type already instantiated, each further call site costs 0.03 ms [0.01, 0.06], about as little as with distinct names.

\subsection{Totals at -O2 (RQ4)}\label{sec:4.5}

With optimization and code generation, several totals are not linear in N (\S{}3.6), so we compare marginal costs at fixed N. At N = 400 native enumeration's marginal cost is 0.61$\times$ [0.60, 0.61] Boost.PFR's (S2: 0.61) and 1.21$\times$ P1061's (S2: 1.22), against 7.96$\times$ P1061's at the front end. Between the inline mechanisms the back end dilutes the front-end difference, because GCC's optimization-and-generation phase takes the same time for every inline arm. Over N = 0 at N = 400 it takes 2.16 s for hand-written code, 2.19 for the X-macro, 2.19 for P1061, and 2.18 and 2.16 s for native enumeration without and with the annotation query. The two callback arms match each other: 3.12 and 3.14 s. Boost.PFR takes 3.97 s and its C++17 path 5.36 s. Equal optimizer time is what one would expect if the inlined code is equivalent; we did not compare the intermediate code. Against Boost.PFR the back end creates a difference instead. Outside the optimizer the two cost about the same, 0.58 and 0.54 s over N = 0 at N = 400, but the optimizer spends 3.97 s on Boost.PFR against 2.18 s on the native arm. About half of that optimizer difference comes with the callback shape, which costs the optimizer the same with P2996 and with P1061 (3.14 and 3.12 s). Behind the same callback API native enumeration costs 0.85$\times$ [0.83, 0.85] Boost.PFR's (S2: 0.84). The callback shape alone multiplies the native arm's marginal cost by 1.38. Most of the inline native arm's advantage at \texttt{-O2} therefore comes from its API shape, not from the enumeration mechanism. The rest is consistent with Boost.PFR's per-field dispatcher (\S{}4.2), one specialization per field index for the optimizer to inline. This is an inference from the source, as we did not inspect the intermediate code.

These ratios move with N, because the optimizer's cost per struct is not constant: in both sessions, the cell at N = 100 of every light arm in FullN except Boost.PFR's C++17 path lies above the line through the arm's other cells. Native / Boost.PFR is 0.68, 0.79, 0.63 and 0.61 at N = 50, 100, 200 and 400 --- below 1 at every N --- and native / P1061 1.19, 1.12, 1.23 and 1.21. Boost.PFR's C++17 path costs 1.73$\times$ its C++26 path at N = 400 (1.39--2.45$\times$ over N = 50--400); the annotation-querying native arm costs 1.31$\times$ hand-written code with distinct names, and one overloaded name costs 1.08$\times$ distinct names. With one function per call site the annotation-querying arm's ratio to Boost.PFR is 0.73$\times$ (against 0.65$\times$ in one function), so the benchmark's layout lowers it, in the native arm's favor, but does not change its direction. For the libraries, Glaze's traditional / P2996 ratio is 1.27, 1.36 and 1.48 at N = 25, 50 and 100, and reflect-cpp's 1.25, 1.50 and 1.64 --- still rising at the end of the sweep, against 2.00 and 2.08 at the front end.

\begin{figure*}[t]
\centering
\includegraphics[width=\textwidth]{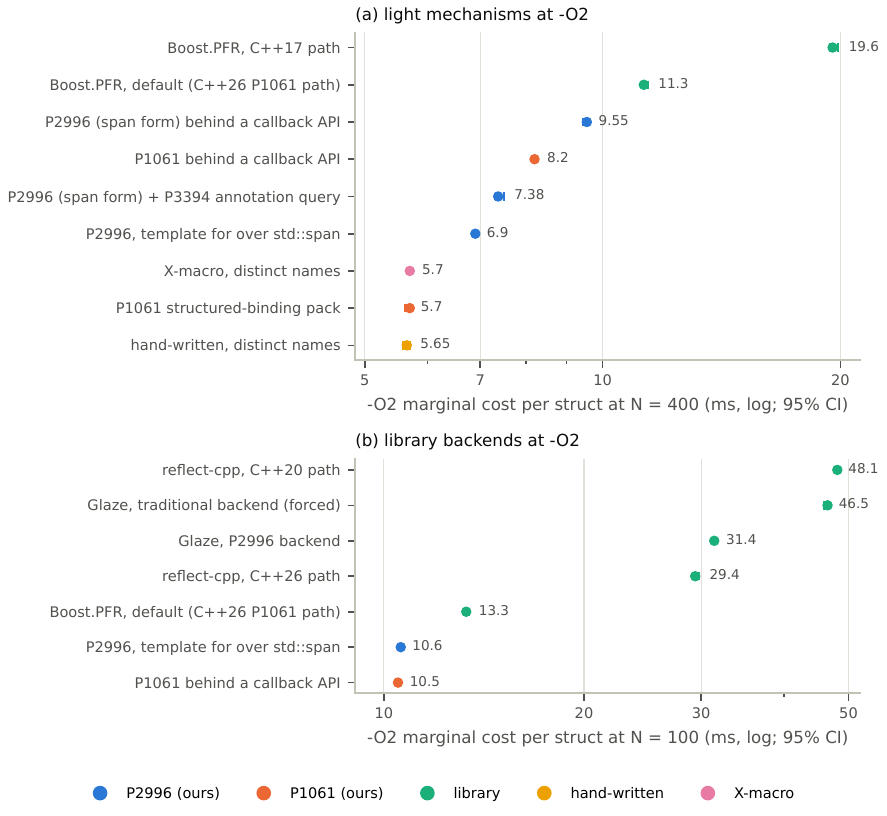}
\Description{Cost per struct at \texttt{-O2}, scalar corpus, M = 16, session S1, log scale: the marginal cost at the largest N divided by N, not a slope, because several \texttt{-O2} curves are not linear (\S{}3.6); whiskers are 95\% bootstrap intervals}
\caption{Cost per struct at \texttt{-O2}, scalar corpus, M = 16, session S1, log scale: the marginal cost at the largest N divided by N, not a slope, because several \texttt{-O2} curves are not linear (\S{}3.6); whiskers are 95\% bootstrap intervals. The P2996 arms use the span form of Listing~\ref{lis:1}. (a) FullN at N = 400: the light arms, including Boost.PFR's C++17 path (\texttt{manual} and the \texttt{*\_\allowbreak{}split} controls omitted). (b) FullHeavy at N = 100: both paths of reflect-cpp and Glaze, with the native arm, Boost.PFR and \texttt{p1061\_\allowbreak{}visit} as references. At N = 100 the light arms cost more per struct than at N = 400 (in FullN their cells at N = 100 lie above the line through their other cells, \S{}4.5), and the libraries' cost per struct still falls with N.}
\label{fig:5}
\end{figure*}

\subsection{Richer corpora, sessions and compiler versions (RQ4)}\label{sec:4.6}

\textbf{Richer corpora.} On strings and containers (rich corpus), native enumeration and Boost.PFR again cost about the same: 0.96$\times$ [0.95, 1.03]. Here each library's C++26 path stays cheaper than its previous path: Glaze by 2.06$\times$, reflect-cpp by 2.20$\times$ and Boost.PFR by 4.91$\times$ (its C++17 path is still close to a line over N $\leq$ 200, \S{}4.3). On the heterogeneous type graph (varied field counts, enums, nesting, vectors and optionals of structs), native enumeration costs 0.99$\times$ Boost.PFR over N $\leq$ 1,600 (HeteroN) and 1.07$\times$ over N $\leq$ 200 (HeteroHeavy). Against P1061 it costs 1.55$\times$ (scalar corpus: 7.96$\times$). This ratio falls because the graph adds a per-struct cost of similar size to every arm. Hand-written code costs 1.98 ms per struct, against 0.23 ms on the scalar corpus; P1061 2.08 against 0.22 ms; native enumeration 3.23 against 1.72 ms. The absolute increment of native enumeration over hand-written code changes little across corpora: 1.49 ms per struct on the scalar corpus, 1.21 ms on the rich corpus and 1.25 ms on the heterogeneous graph. Yet the ratio to hand-written code falls from 7.46$\times$ to 5.17$\times$ and 1.64$\times$. The libraries' C++26 paths stay cheaper on this graph, by different margins: Glaze's backend ratio is 1.82, reflect-cpp's only 1.21$\times$ [1.20, 1.22]. Boost.PFR's C++17 path costs 4.37$\times$ its C++26 path at N = 1,600. On this graph reflect-cpp's C++26 path costs 77.09 ms per struct, against 14.66 ms on the scalar corpus (Glaze's P2996 backend: 11.17 ms). We did not isolate which feature of the graph drives reflect-cpp's cost.

\textbf{GCC 16.1.0 against 16.2.0} (paired, same rounds). GCC 16.2.0 is faster on every arm that both versions compiled with \texttt{-freflection}, but by different factors, so comparisons between mechanisms depend on the version. In CompilerN (N $\leq$ 2,000), GCC 16.1.0 costs 1.124$\times$ GCC 16.2.0 per struct on the native arm, 1.995$\times$ on P1061, 1.550$\times$ on Boost.PFR and 1.748$\times$ on hand-written code. On the libraries (CompilerHeavy, N $\leq$ 200) the difference is smaller: 1.102$\times$ on Glaze's P2996 backend, 1.088$\times$ on its traditional backend, 1.065$\times$ on reflect-cpp's C++26 path and 1.036$\times$ on its C++20 path. In CompilerHeavy, where the libraries and the native arm share rounds, the native arm's ratio is 1.247$\times$ [1.212, 1.304], above every library's, against 1.124$\times$ (above) in CompilerN, so the native arm's version ratio depends on the sweep. In CompilerN, native enumeration costs 0.71$\times$ Boost.PFR, 4.16$\times$ P1061 and 4.44$\times$ hand-written code per struct on GCC 16.1.0, against 0.99$\times$, 7.39$\times$ and 6.91$\times$ on GCC 16.2.0. The annotation factor (1.33$\times$ and 1.33$\times$) and the library backend A/Bs barely move: Glaze's ratio is 1.93 on GCC 16.1.0 and 1.95 on 16.2.0, reflect-cpp's 1.97 and 2.03. For Boost.PFR and P1061 the version difference appears only with \texttt{-freflection}. Compiled without it, the two versions cost the same per struct (GCC 16.1.0 / 16.2.0: 1.000 and 0.990). In the same rounds the flag costs 1.703 on Boost.PFR and 2.578 on P1061 with GCC 16.1.0, against 1.092 and 1.264 with GCC 16.2.0. Hand-written code was not compiled without the flag on GCC 16.1.0. We did not identify which change in GCC 16.2.0 is responsible. GCC's C++ front-end ChangeLog lists two changes to the check for consteval-only types between the releases, PR125179 and PR125280~\cite{gccpr125280}.

\textbf{Sessions.} S2 ran the same plan with the same compilers and code on a different day. For the per-type slopes behind the headline numbers the S2/S1 ratio lies between 0.972 (P1061 on GCC 16.1.0, CompilerFlag) and 1.020 (Boost.PFR, FlagCost); in S2 the flag costs P1061 2.48$\times$ on GCC 16.1.0, against 2.578$\times$ in S1. Across all 102 per-arm slopes of the plan that pass the linearity check and rise by at least ten timer steps (the 96 of \S{}3.5 and the six external wall-time slopes of ReportOverhead), the median difference between the sessions is 0.7\%, 90\% differ by at most 2.5\%, and the largest is 8.8\% (the native arm's per-field slope in the field sweep, 1.088 [1.066, 1.140]; no headline number uses that slope). The within-session intervals do not cover the difference between the sessions: 11 of those 102 S2/S1 ratios have an interval that excludes 1, where about five would if the intervals covered it. Two sessions cannot tell whether this is variation between days or under-coverage of percentile intervals from eleven rounds; either way, differences of a few percent between arms should be read against both sessions, not against one interval. The headlines that are not slopes reproduce as well: in S2, Boost.PFR's C++17 / C++26 ratio is 4.5 at N = 500 and 12.8 at N = 4,000, and Glaze's ratio at M = 128 is 0.70. Native enumeration's ratio to Boost.PFR at M = 128 does not reproduce (0.80 in S1, 0.89 in S2), and we do not use it. Pooled over both sessions, the main ratios are 7.88 (native / P1061), 1.03 (native / Boost.PFR), 1.37 (the annotation query), 1.99 (Glaze) and 2.08 (reflect-cpp).

\subsection{Maintainability (RQ5)}\label{sec:4.7}

Per-type source is an analytic count, not a measurement. With 16 fields a hand-written serializer is 19 lines per type and an X-macro field list 18 (one field per line): both name every field once per type, so both grow as O(N$\cdot$M), while the P2996, P1061 and library arms need no per-type source. Their one-off generic code is 28 lines for the native serializer (42 with the annotation query and its tag type), 21 for P1061, and 21 for each library adapter, excluding the libraries themselves; the X-macro's shared code is 25 lines.

\section{Related Work}\label{sec:5}

\textbf{Native reflection alone.} Romeo~\cite{romeo2026a} times native P2996 on GCC 16 (prerelease, hyperfine, single translation units compiled and linked at \texttt{-O0}): including \texttt{\textless{}meta\textgreater{}} costs about 155 ms on a release build, and each reflected type 1.1--1.7 ms. His first numbers came from an assertion-enabled compiler; the release rebuild was about 20--50\% faster depending on the row. His follow-up~\cite{romeo2026b} compares reflection with an enum library and X-macros for enum-to-string on GCC 16.1.1, where \texttt{\textless{}meta\textgreater{}} dominates. Neither compares library or hand-written field enumeration, and both include the fixed \texttt{\textless{}meta\textgreater{}} cost in the per-TU totals; our per-type numbers exclude it by construction, and our 30 ms for \texttt{\textless{}meta\textgreater{}} (\S{}4.1) is its cost over a translation unit that already includes \texttt{\textless{}string\textgreater{}}, \texttt{\textless{}vector\textgreater{}} and \texttt{\textless{}optional\textgreater{}}, not over an empty one.

\textbf{Concurrent measurements.} The field-reflection library~\cite{matsuda2026} (table in its README since 2026-10-12) switches to a \texttt{std::\allowbreak{}meta} backend on GCC 16 with \texttt{-std=\allowbreak{}c++26 -freflection} and reports the compile time added per 8-field struct visited with names (slope between 10 and 200 types in one translation unit, \texttt{-c -O0}, best of 3): 20 ms for its C++26 backend against 18 ms for Boost.PFR's \texttt{for\_\allowbreak{}each\_\allowbreak{}field\_\allowbreak{}with\_\allowbreak{}name} on GCC 16, and 23 ms for its C++20 backend against 22 ms for Boost.PFR on GCC 15. Its two backends are timed on different compilers, without intervals, P1061, hand-written or macro baselines. mp-units~\cite{pusz2026} reports about $-$10\% to $-$12\% for its reflection-based conversion engine on one workload and +8\% to +33\% for its value-domain expression operations on one ISQ-heavy translation unit (near-neutral on lean ones), measured on WSL only, with the author cautioning that ``the sign is trustworthy, the magnitude is not''. reflect-moc's measurements of a moc replacement~\cite{reflectmoc} find totals between about 12\% below and 6\% above moc's, on a loaded machine. simdjson reports a 6\% slowdown with reflection enabled (CppCon 2025~\cite{lemire2025cppcon}, clang-p2996) and later cut its reflection example's build by 35\% by changing what every TU instantiates~\cite{simdjsonpr2894}. Kolpackov (GCC PR125329~\cite{gccpr125329}) measures \texttt{\textless{}meta\textgreater{}} across 128 TUs with and without header units. Matek~\cite{matek2025} finds reflection-based type identity about 4$\times$ faster than template-based identity on a Clang fork. GCC's own reports track the cost of the flag and of reflection-heavy constant evaluation: PR124925~\cite{gccpr124925} (19\% on \texttt{\textless{}bits/\allowbreak{}stdc++.\allowbreak{}h\textgreater{}} in GCC 16.0, later closed as fixed), PR125179~\cite{gccpr125179} (a 40-fold slowdown reduced to 1.1$\times$, in GCC 16.2.0), and Sajdak (PR128900--PR128902~\cite{gccpr128900,gccpr128901,gccpr128902}, November 2026), who counts the front end's instructions on 16.2.0: \texttt{identifier\_\allowbreak{}of} costs about 35 k instructions per character of the name, because the returned \texttt{string\_\allowbreak{}view} is measured by a \texttt{char\_\allowbreak{}traits} loop. reflect-cpp's maintainer found~\cite{reflectcpppr537}, with Clang's \texttt{-ftime-trace}, that serialization code generation rather than reflection dominates his build; our library arms time field visitors, not serializers, so they do not test this.

\textbf{Libraries.} Glaze documents its P2996 backend as new capabilities; we found no published compile-time A/B of it. The one Glaze compile-time datum we found (\texttt{make\_\allowbreak{}keys\_\allowbreak{}info\textless{}20\textgreater{}}, 45.1 $\rightarrow$ 26.0 ms in the v7.7.0 release notes~\cite{glazev770}) is unrelated to the P2996 backend and was measured with \texttt{-ftime-trace}; after v9.1.0, Glaze's main branch changed how its P2996 backend enumerates members (PR \#3112~\cite{glazepr3112}). reflect-cpp's documentation claims its C++26 path is cheaper to compile without data~\cite{reflectcppdocs}, and its CI does not build that path. Boost.PFR ships an experimental P2996 engine (since Boost 1.91.0) that its \texttt{for\_\allowbreak{}each\_\allowbreak{}field} does not use under \texttt{-std=\allowbreak{}c++26}. Glaze PR \#2848~\cite{glazepr2848} (merged 2026-08-29) replaced its tuple implementation for licensing reasons and with it an N-candidate \texttt{operator[]} overload set behind \texttt{get}; it reports $-$41\% compile time on a tuple microbenchmark and $-$1\% to $-$3\% on a whole \texttt{glaze.\allowbreak{}hpp} translation unit (Apple clang 21). That is the closest practitioner evidence we found for the overload-set effect of \S{}4.4, which we measure across the customization idioms of six serialization and formatting libraries, with call sites and candidates varied separately.

\textbf{Adopted, not measured.} Talks and libraries that use P2996 for serialization or code generation report capability, not compile cost: Webber~\cite{webber2026}, Piper~\cite{piper2026}, Chen~\cite{chen2026}, simdjson 5.0~\cite{lemire2026simdjson5}.

\textbf{Measuring compile time.} Metabench~\cite{metabench} isolates a code block by subtracting a baseline compile, as our marginal costs do; Chochlík (P2560R0~\cite{wg21p2560}) compared, for SG7, the compile cost of value-based and type-based reflection APIs by emulating up to a million metaobjects in generated sources and subtracting a baseline compile, before any P2996 implementation existed; ctbench~\cite{ctbench2023} measures scaling with Clang's \texttt{-ftime-trace} and notes that there is ``no clear cut methodology'' to compare metaprogramming strategies; Templight~\cite{porkolab2006}, a debugger of template instantiation, was later used to profile it~\cite{porkolab2009}. Gregor and Järvi~\cite{gregor2008} measured a new language feature, variadic templates, against its preprocessor emulation on GCC, broken down by phase --- the closest precedent to this study. Our protocol follows the benchmarking literature: randomized interleaved trials~\cite{abedi2017}, awareness of setup bias~\cite{mytkowicz2009}, effect sizes with confidence intervals~\cite{kalibera2013}, the number of rounds checked by resampling (\S{}3.6), bootstrap intervals~\cite{efron1993,barrett2017}, and documented factors~\cite{hoefler2015}.

\section{Threats to Validity}\label{sec:6}

\textbf{Construct validity.} \emph{One native form.} S3 times three other formulations of the member query (\S{}4.2), but only at the front end, on the scalar corpus with 16 fields, on GCC 16.2.0 and in one session. Every native number from the rich and heterogeneous corpora, the field sweep, \texttt{-O2} and GCC 16.1.0, and every number of the annotation and callback arms (\texttt{reflect}, \texttt{reflect\_\allowbreak{}visit}), uses the span form and so includes the cost of iterating over the span; S3's front-end contrast suggests that, on GCC 16.2.0, these front-end numbers would be lower over a constant array, which we did not test.

\emph{Instrument and flag.} The \texttt{*\_\allowbreak{}noreport} and \texttt{*\_\allowbreak{}noflag} controls measure the cost of the two options that the other arms use, \texttt{-ftime-report} and \texttt{-freflection} respectively (\S{}3.1, \S{}4.1). Without the report, the native arm's ratio to hand-written code would be about 9\% larger on external wall time, while its ratio to Boost.PFR would not change (\S{}4.1); we did not measure the report's overhead for P1061, the X-macro or the other arms, so its effect on the headline ratio to P1061 is unknown. Because only the phase timers partition a compile, and of the other timers only the two stack timers exclude each other (\S{}4.2), we report timer shares one at a time and add up only the two stack timers. The flag, which GCC labels experimental, also slows the arms that do not use reflection. Compared with those arms compiled without the flag, as in a project that does not enable reflection, the native arm's per-struct ratios would be larger by about the flag's factor for each arm (\S{}4.1).

\textbf{Internal and conclusion validity.} \emph{Confounds.} The code-identical control shows that what a translation unit parses can change per-type cost by a few percent (\S{}4.2), so differences of that size between the native arms, which parse \texttt{\textless{}meta\textgreater{}}, and the P1061, hand-written and Boost.PFR arms cannot be attributed to the mechanism. We did not test whether \texttt{\textless{}meta\textgreater{}} alone has this effect, nor did we test for setup bias (\S{}5) from the one extra \texttt{-D} on \texttt{pfr\_\allowbreak{}native}'s command line. If \texttt{\textless{}meta\textgreater{}} has it, the native ratios to those arms are overstated by up to a few percent: enough to matter for the span form's parity with Boost.PFR, not for the larger ratios. Attributing the span/array gap mainly to the iterating expansion, rather than to the untimed calls that \texttt{define\_\allowbreak{}static\_\allowbreak{}array} adds, is an inference (\S{}4.2); contribution 2 and \S{}7 state it as one.

\emph{A session added after the main analysis.} S3 was added after the results of S1 and S2 had been analyzed and reviewed (\S{}3.5). Its analysis rules fix which contrast may carry an attribution, not the outcome. They were set before the session, but after six of its seven arms (all but the array form, added later) had each been compiled once informally and, with the array form added, a one-round smoke test of the whole job had run. Neither is used in the analysis, but the smoke test already pointed to the span/array difference. The shared arms reproduce S1, and the span/array factor of two far exceeds the variation between S1 and S2 (\S{}4.2), but because S3 is a single session, its intervals understate day-to-day variation, and every array-form result rests on it alone.

\emph{Machine and sessions.} A job's arms share rounds, in random order between compiles of a drift sentinel, on the performance cores at high priority (\S{}3.4, \S{}3.5), so load and drift add noise rather than, in expectation, a bias between arms. Dropping the compiles above 15\% load changes the slopes little, and in S1 and S2 long compiles spent almost no time off the CPU (S3 somewhat more; \S{}3.5). Still, wall time inside the compiler is load-sensitive, the machine was an ordinary desktop with its usual applications open and Microsoft Defender's real-time protection on (\S{}3.7), and frequency changes cannot be ruled out: the power plan was fixed (High performance) and recorded (\S{}3.5), but core frequency was not recorded. Within-session intervals do not cover all variation between S1 and S2 (\S{}4.6).

\textbf{External validity.} \emph{One compiler family.} Key arms were also timed on GCC 16.1.0 (\S{}4.6), and Boost.PFR's C++17 path probed on GCC 15.2.0, where its quadratic term appears as well (\S{}4.3); we did not test Clang. As of November 2026, GCC 16 is the only released compiler with P2996 that emits object code: Clang's upstream support covers primitive types only~\cite{llvmpr190356}; Bloomberg's clang-p2996~\cite{clangp2996} is a self-described experimental fork; EDG 7.0~\cite{edg70} (open source since September 2026) runs P2996 code through a C-generating back end, so its timings would include a C compile, and on Compiler Explorer it rejected a probe with the expansion statement (\texttt{template for}, P1306) that our native arms use. Every number is thus a property of GCC 16's implementation and can move between point releases: from GCC 16.1.0 to 16.2.0 the mechanism ratios moved far more than the library backend A/Bs (\S{}4.6). Without \texttt{-freflection}, P1061 and Boost.PFR cost the same on GCC 16.1.0 and 16.2.0, so their version difference lies in the flag, consistent with GCC 16.2.0's \texttt{-freflection} fixes (PR125179 and PR125280 among them), which we did not bisect; for the native arm, which needs the flag, the causes cannot be separated.

\emph{Standard and implementation.} The standard specifies that the expansion statement of the span form iterates element by element and that of the array form destructures the array (an expression of array type is never expansion-iterable); what each costs is a property of GCC 16.2.0 and of libstdc++'s \texttt{define\_\allowbreak{}static\_\allowbreak{}array}, which another compiler, library or release need not share. GCC 16.2.0 also declares the span form's artificial variables \texttt{static}, a specifier that CWG3044~\cite{cwg3044} removed from the specification; this exposes the span form to the quadratic registration of block-scope statics (\S{}4.2, PR128928~\cite{gccpr128928}). A patch posted to gcc-patches on 2026-11-16~\cite{jelinek2026}, not merged by 2026-11-21, drops those statics for C++26. We expect the patch to remove that part of the span form's cost but did not measure this.

\emph{Corpora and layout.} The corpora sample a range of type shapes (\S{}3.2) but neither reproduce nor bound real codebases: no corpus contains base classes, user-defined class templates or private members, and field names have two to four characters. The native arms query no names, but the C++26 paths of Glaze and reflect-cpp call \texttt{identifier\_\allowbreak{}of}, whose cost on GCC 16.2.0 grows with the length of the name (PR128900~\cite{gccpr128900}), for every field, and their pre-C++26 paths parse names out of \texttt{\_\allowbreak{}\_\allowbreak{}PRETTY\_\allowbreak{}FUNCTION\_\allowbreak{}\_\allowbreak{}}; longer names could change the library arms' costs, and possibly their backend ratios, but not the native arms'. At \texttt{-O2}, serializing every struct from one function favors the annotation arm against Boost.PFR, though not enough to change the direction (\S{}4.5); the arms behind the \texttt{-O2} ratios of \S{}7 (\texttt{reflect\_\allowbreak{}plain}, \texttt{reflect\_\allowbreak{}visit}, \texttt{p1061}) have no layout control.

\emph{Build, host and hardware.} Both compilers were built alike (\S{}3.7): non-bootstrapped, with release checking, by the MSYS2 host GCC 15.2.0, with configure's default flags (\texttt{-g -O2} in the GCC 16.2.0 build log). A three-stage bootstrap (GCC's default), or the profile feedback or LTO that distributions may add, can make a compiler faster, not necessarily by the same factor in every phase, so absolute times may differ, and so may the ratios between arms whose time falls in different phases. The front end is the same code on every host, but on this Windows host (MinGW-w64) configure finds no anonymous \texttt{mmap} (checked for the GCC 16.2.0 build), so GCC's garbage collector takes its pages from \texttt{malloc} page groups rather than from \texttt{mmap}. That, like another microarchitecture or memory system, could shift the ratios between arms whose working sets differ several-fold (\S{}4.2, \S{}4.3). We did not repeat the measurements with a bootstrapped compiler, on Linux or on another machine.

\textbf{Scope.} The library arms time each library's field visitor (\S{}3.1), not its serializer, which does more work per type (keys, formatting). Our native and P1061 arms compute no field names (the binary format needs none; P1061 cannot provide them), while Glaze's and reflect-cpp's visitors do (\S{}4.3), so the native arms' ratios to Glaze and reflect-cpp are not like-for-like in this respect and favor the native arms (Boost.PFR's \texttt{for\_\allowbreak{}each\_\allowbreak{}field} computes no names); a name-keyed format such as JSON would add \texttt{identifier\_\allowbreak{}of} to a native serializer, at a cost we did not measure. Deserialization is neither implemented nor measured. Two further forms were not timed: a native member list recomputed per index, as in Glaze's P2996 backend at v9.1.0 (\S{}4.3), and an expansion statement that destructures the struct itself, without reflection (\S{}2).

\section{Conclusion}\label{sec:7}

On GCC 16.2.0, native P2996 enumeration of a struct with 16 scalar fields costs the front end between 0.82 and 2.02 ms per struct across the four formulations timed in session S3. The form used in several of P2996R13's own examples is an expansion statement over the \texttt{std::\allowbreak{}span} that \texttt{define\_\allowbreak{}static\_\allowbreak{}array} returns. It costs 1.72 ms per struct over N $\leq$ 4,000 (S3: 1.73), 7.96$\times$ as much as a P1061 structured-binding pack, which itself costs about as much as a hand-written serializer with its own name per type. About half of the span form's cost comes with the span, which we attribute mainly to GCC's evaluation of the iterating expansion statement that C++26 specifies over it (\S{}4.2). The same loop over a constant array costs 0.47$\times$ as much. Written inline, the span form costs about as much as Boost.PFR's default \texttt{for\_\allowbreak{}each\_\allowbreak{}field}, and behind the same callback API 1.21$\times$ as much. Over a constant array, inline native enumeration costs 0.49$\times$ Boost.PFR's \texttt{for\_\allowbreak{}each\_\allowbreak{}field}; no callback-shaped array arm was timed. A P3394 annotation query on every member multiplies the span form's cost by 1.37. The span form adds 1.21--1.49 ms of front-end work per struct over hand-written code on our three corpora. Thus its ratios depend on everything else a struct costs: 1.55$\times$ P1061 on a heterogeneous type graph, and 1.21$\times$ at \texttt{-O2}, where the optimizer spends about the same time on every inline mechanism. At \texttt{-O2} the inline native serializer costs 0.61$\times$ Boost.PFR's, mostly because a callback API costs the optimizer more. Behind the same callback API the ratio is 0.85.

\textbf{What follows for practice, on GCC 16.2.0.} In a whole translation unit the choice of library outweighs the choice of enumeration mechanism. With 200 structs of 16 scalar fields the front end takes 0.59 s with inline native enumeration and 0.56 s with Boost.PFR. The C++26 paths of reflect-cpp and Glaze, whose visitors also compute field names, take 3.86 and 4.13 s (Table~\ref{tab:3}). For an inline native serializer, an expansion statement over a constant array, \texttt{[:std::\allowbreak{}meta::\allowbreak{}reflect\_\allowbreak{}constant\_\allowbreak{}array(\allowbreak{}...):]} (standard C++26), is the cheaper choice at the front end (0.47$\times$ the span form's cost per struct of 16 scalar fields, S3): with fixed costs included, a whole translation unit takes 0.61$\times$ the span form's front-end time with 1,000 structs and 0.50$\times$ with 4,000. We did not time the array form on the other corpora or at \texttt{-O2}; at \texttt{-O2} the span form already costs only 1.21$\times$ P1061 at 400 structs (\S{}4.5), so the array form's front-end saving would likely matter less there. Code that needs neither names nor annotations, such as a binary format, hashing or comparison, can enumerate with a P1061 pack the fields of a struct that structured bindings can decompose. For our serializer this costs about as much as hand-written code, with no per-type source. We did not time a further standard form, an expansion statement that destructures the struct. A library whose users customize serialization by overloading one name in one namespace makes every resolution of that name consider all of its overloads in that namespace, so the cost grows faster than linearly with the number of types (\S{}4.4). Hidden friends, member functions and class-template specializations avoid the shared set (Table~\ref{tab:4}). Every per-type number excludes fixed costs, which precompiled headers or modules may amortize; we measured neither. Per-type costs recur in every translation unit that serializes the type.

Per type, each library's C++26 path is cheaper to compile than its pre-C++26 path at 16 fields. Glaze's gain is 2.00$\times$, reflect-cpp's 2.08$\times$, and Boost.PFR's 4.4$\times$ at 500 types and 12.8$\times$ at 4,000, because its C++17 path grows almost quadratically with the number of types. But the gain is not a constant of the library. Glaze's is reversed at 128 fields (0.70$\times$, pinned release v9.1.0); reflect-cpp's falls to 1.21$\times$ on a heterogeneous type graph. In a whole translation unit the headers dilute the gain. Hand-written serializers that share one name grow by 3.5--4.0$\times$ per doubling of the number of types from 500 to 1,000 (K = C = N) for the five idioms of Table~\ref{tab:4} that share one lookup set. The hand-written arm with one overloaded name grows faster still from 2,000 to 4,000 types (\S{}4.4) and overtakes native reflection in the span form between 1,000 and 2,000 types.

The toolchain also moves the comparisons. The \texttt{-freflection} flag makes code that does not use reflection 1.017--1.264$\times$ costlier per type on GCC 16.2.0 and up to 2.578$\times$ on GCC 16.1.0, and between these two point releases inline native enumeration in the span form moves from 0.71$\times$ to 0.99$\times$ Boost.PFR's cost, while the library backend A/Bs barely change. Comparisons between mechanisms are therefore properties of a compiler release and of the formulation measured, and should be repeated as GCC's reflection support matures and other compilers ship it.

\section*{Artifact}

The artifact is archived at \url{https://doi.org/10.5281/zenodo.24443567} (the version this paper reports) and developed at \url{https://github.com/DenysKachalenko/CPP26FieldEnumerationCompileCostArtifact}; its code is released under the MIT license, its data and text under CC BY 4.0. \texttt{DATA.\allowbreak{}md} documents every file and every column of the raw data. Every number and figure can be regenerated from the archived raw data without a compiler, with Python, NumPy and Matplotlib (3.13.1, 2.5.1 and 3.11.0 here; \texttt{requirements.\allowbreak{}txt}), run from \texttt{Harness/\allowbreak{}}:

\begin{itemize}
\item \texttt{python papernumbers.\allowbreak{}py .\allowbreak{}.\allowbreak{}/\allowbreak{}Data/\allowbreak{}Sessions/\allowbreak{}S1 .\allowbreak{}.\allowbreak{}/\allowbreak{}Data/\allowbreak{}Sessions/\allowbreak{}S2 .\allowbreak{}.\allowbreak{}/\allowbreak{}Data/\allowbreak{}Sessions/\allowbreak{}S3} writes \texttt{Data/\allowbreak{}PaperNumbers.\allowbreak{}json};
\item \texttt{python paperlint.\allowbreak{}py} checks every tagged number in \texttt{Paper.\allowbreak{}md} against it;
\item \texttt{python figures.\allowbreak{}py .\allowbreak{}.\allowbreak{}/\allowbreak{}Data/\allowbreak{}Sessions/\allowbreak{}S1 .\allowbreak{}.\allowbreak{}/\allowbreak{}Data/\allowbreak{}Sessions/\allowbreak{}S3} draws \texttt{Figures/\allowbreak{}}.
\end{itemize}

Re-measuring needs both compilers, built with \texttt{Harness/\allowbreak{}build\_\allowbreak{}gcc.\allowbreak{}sh}, and one unattended run per session (\texttt{Harness/\allowbreak{}overnight.\allowbreak{}sh}, \texttt{RunBook.\allowbreak{}md}); the compiles of S1 and S2 took about 5.7 hours each, those of S3 about 16 minutes. The archive contains the generator (\texttt{Harness/\allowbreak{}generate.\allowbreak{}py}, \texttt{Harness/\allowbreak{}overload.\allowbreak{}py}), the arm registry and manipulation checks (\texttt{Harness/\allowbreak{}arms.\allowbreak{}py}, \texttt{Harness/\allowbreak{}preflight.\allowbreak{}py}), the harness (\texttt{Harness/\allowbreak{}campaign.\allowbreak{}py}, \texttt{Harness/\allowbreak{}compiler.\allowbreak{}py}, \texttt{Harness/\allowbreak{}machine.\allowbreak{}py}, \texttt{Harness/\allowbreak{}plan.\allowbreak{}py}), the unattended run (\texttt{Harness/\allowbreak{}overnight.\allowbreak{}sh}), the Boost.PFR probe (\texttt{Harness/\allowbreak{}pfrprobe.\allowbreak{}py}, \texttt{Data/\allowbreak{}PfrProbe.\allowbreak{}json}), statistics and paper numbers (\texttt{Harness/\allowbreak{}stats.\allowbreak{}py}, \texttt{Harness/\allowbreak{}papernumbers.\allowbreak{}py}, \texttt{Harness/\allowbreak{}paperlint.\allowbreak{}py}, \texttt{Harness/\allowbreak{}figures.\allowbreak{}py}, \texttt{Data/\allowbreak{}PaperNumbers.\allowbreak{}json}, \texttt{Figures/\allowbreak{}}), the GCC build script (\texttt{Harness/\allowbreak{}build\_\allowbreak{}gcc.\allowbreak{}sh}), pins (\texttt{Deps.\allowbreak{}json}), CMake tests, the reproducers of PR128928 (\texttt{Repro/\allowbreak{}}), raw per-run data, environment records and a copy of the measured code per session (\texttt{Data/\allowbreak{}Sessions/\allowbreak{}S1}, \texttt{S2}, \texttt{S3}), and the run book (\texttt{RunBook.\allowbreak{}md}).

\begin{acks}
The author conceived the study, made its design and methodological decisions, ran the measurements and checked the results. An AI assistant (Claude, by Anthropic) was used under the author's direction to write the measurement harness and the analysis scripts, to draft and edit the text, to search the literature and to review the paper. The harness is not trusted on its own: a value oracle with a negative control checks every arm's output against a hand-written reference, the manipulation checks of \S{}3.3 check every single-factor pair, and the full plan was measured in two sessions. Every measured number in the paper is generated by the analysis scripts from the archived raw data and checked against the text automatically (Artifact). The author takes full responsibility for the content.
\end{acks}

\bibliographystyle{ACM-Reference-Format}
\bibliography{References}

\end{document}